\documentclass[11pt,a4paper]{article}
\pdfoutput=1
\usepackage{jheppub}

\usepackage{graphicx}
\usepackage{amsmath}
\usepackage{amssymb}
\usepackage{braket}
\usepackage{hyperref}
\usepackage{frontespizio}
\usepackage{comment}
\usepackage{epsfig}
\usepackage{float}
\usepackage{multirow}
\usepackage{color}
\usepackage[font=small,skip=5pt]{caption}

\usepackage{subfigure}
\newcommand{\bea}{\begin{eqnarray}}
\newcommand{\eea}{\end{eqnarray}}
\newcommand{\be}{\begin{equation}}
\newcommand{\ee}{\end{equation}}
\newcommand{\ba}{\begin{align}}
\newcommand{\ea}{\end{align}}

\newcommand{\Kahler}{\ensuremath{\text{K}\ddot{\text{a}}\text{hler}\,}}

\title{The Fate of Dense Scalar Stars}
\author[1]{Francesco Muia}
\author[2,3]{, Michele Cicoli}
\author[4]{, Katy Clough}
\author[2,3]{, Francisco Pedro}
\author[4,5]{, Fernando Quevedo}
\author[3]{, Gian Paolo Vacca}

\affiliation[1]{\small \it ICTP, Strada Costiera 11, Trieste 34014, Italy}
\affiliation[2]{\small \it Dipartimento di Fisica e Astronomia, Universit\'a di Bologna, via Irnerio 46, 40126 Bologna, Italy}
\affiliation[3]{\small \it INFN, Sezione di Bologna, viale Berti Pichat 6/2, 40127 Bologna, Italy}
\affiliation[4]{\small \it Astrophysics, University of Oxford, DWB, Keble Road, Oxford OX1 3RH, UK}
\affiliation[5]{\small \it DAMTP, Centre for Mathematical Sciences, Wilberforce Road, Cambridge, CB3 0WA, UK}
\emailAdd{fmuia@ictp.it}
\emailAdd{michele.cicoli@unibo.it}
\emailAdd{katy.clough@physics.ox.ac.uk}
\emailAdd{francisco.pedro@bo.infn.it}
\emailAdd{fq201@damtp.cam.ac.uk}
\emailAdd{Gianpaolo.Vacca@bo.infn.it}
\abstract{

Long-lived pseudo-solitonic objects, known as \textit{oscillons/oscillatons}, which we collectively call \textit{real scalar stars}, are ubiquitous in early Universe cosmology of scalar field theories. Typical examples are \textit{axions stars} and \textit{moduli stars}. Using numerical simulations in full general relativity to include the effects of gravity, we study the fate of real scalar stars and find that depending on the scalar potential they are either meta-stable or collapse to black holes. In particular we find that for KKLT potentials the configurations are meta-stable despite the asymmetry of the potential, consistently with the results from lattice simulations that do not include gravitational effects. For $\alpha$-attractor potentials collapse to black holes is possible in a region of the parameter space where scalar stars would instead seem to be meta-stable or even disperse without including gravity. Each case gives rise to different cosmological implications which may affect the stochastic spectrum of gravitational waves.}
\begin{document} 
\maketitle

\section{Introduction}
\label{sec:Introduction}

The discovery of the Higgs field~\cite{Aad:2012tfa, Chatrchyan:2012xdj} represents the first observation of a fundamental scalar field\footnote{Even though the possibility that the Higgs is not a fundamental particle is not ruled out yet, see~\cite{Csaki:2015hcd} for a recent review.}. However, there are many reasons to believe that the Higgs is not the only fundamental scalar degree of freedom existing in nature. In fact, most extensions of the Standard Model of particle physics (SM) provide for the existence of scalar fields at various energy scales. Some examples include:

\begin{itemize}

\item{} The existence of a pseudo-scalar field, the QCD axion, is implied by one of the best motivated solutions to the strong CP problem of QCD, via the Peccei-Quinn mechanism~\cite{Peccei:1977hh}, see~\cite{Marsh:2015xka} for a review. 

\item{} At least one scalar field is required in the simplest models of inflation, which is the currently most accepted extension of $\Lambda$CDM: it describes an accelerated expansion of the Universe during its early stages due to the potential energy stored in a scalar field that is undergoing a slow-roll motion, see e.g.~\cite{Baumann:2009ds} for a review. 

\item{} Any supersymmetric extension of the SM includes scalar fields with masses above the electroweak scale, see e.g.~\cite{Martin:1997ns, Weinberg:2000cr}. 

\item{} String theory predicts the existence of several gravitationally coupled scalar fields, called \textit{moduli}, whose vacuum expectation values parametrize the size and shape of the extra-dimensions required for the consistency of the theory~\cite{Ibanez:2012zz}. String theory also predicts the existence of a plethora of axion-like particles, not necessarily related to the strong CP problem of QCD~\cite{Arvanitaki:2009fg, Cicoli:2012sz}.

\item Very recently four-dimensional extensions of the SM (not including gravity) modelled as fundamental QFT have been considered~\cite{Litim:2014uca}. They are based on the UV completion mechanism called \textit{asymptotic safety}, characterized by the presence of a non trivial UV fixed point of the renormalization group which allows to overcome problems such as the presence of Landau poles in the theory. Perturbative analyses of interacting non-Abelian gauge fields, fermions and scalars in the Veneziano limit for different gauge groups and matter representations have shown that generically several fundamental scalar fields are required for the UV fixed point to exist.
\end{itemize}

The classical equations of motion of scalar fields admit (meta-)stable localized solutions\footnote{The energy density associated to a (pseudo-)solitonic solution goes rapidly to zero far away from the centre of the (pseudo-)soliton. While solitons are classically stable due to some conservation law, pseudo-solitons can be long-lived but eventually decay.}, also known as \textit{(pseudo-)solitons}~\cite{Weinberg:1996kr, Rubakov:2002fi, Manton:2004tk, Tong:2005un}. The rapid dispersion of such localized solutions is avoided due to the non-linearity of the equations of motion, giving rise to stable or long-lived compact objects that, if formed during the early Universe, can leave observable signatures in the form of a stochastic background of Gravitational Waves (GWs). The detection of such GW background would provide valuable information about the pre-Big Bang Nucleosynthesis (BBN) era~\cite{Kusenko:2008zm, Dolgov:2011cq, Amin:2011hj, Amin:2014eta, Amin:2013ika, Amin:2010jq, Antusch:2016con, Antusch:2017flz}. In some scenarios, long-lived (pseudo-)solitons can survive till the current time, either providing a natural candidate for dark matter~\cite{Kusenko:1997ad, Kusenko:1997si, Kusenko:2001vu, Olle:2019kbo} or a source of distinct signatures if a fraction of the dark matter is in the form of compact objects, see e.g.~\cite{Bai:2016wpg, Braaten:2016dlp, Levkov:2016rkk, Eby:2017xaw, Hui:2016ltb, Desjacques:2017fmf}.

There exist a plethora of distinct (pseudo-)solitonic compact objects, whose differences can be traced back to the mechanism that guarantees their (meta-)stability, see~\cite{Krippendorf:2018tei} for a recent review. If this is ensured by the conservation of a topological charge, the resulting compact object is a \textit{topological soliton}~\cite{Manton:2004tk}. Examples of this class are kinks, vortices and skyrmions. If the stability is given by the conservation of a Noether charge, then the compact object is a \textit{non-topological soliton}~\cite{Lee:1991ax}: some of the simplest representatives of this class are summarized in Tab.\ \ref{tab:Classification}. We collectively refer to the objects appearing in Tab.\ \ref{tab:Classification}, i.e. non-topological solitons composed by a single scalar field, as \textit{scalar stars}.

A complex scalar field whose Lagrangian obeys a global $U(1)$ symmetry\footnote{E.g. if the complex field $\phi$ has canonical kinetic terms and scalar potential $V = \frac{1}{2} m^2 |\phi|^2 + \frac{g}{4!} |\phi|^4$.} \label{foo:Potential} gives rise to non-topological solitons called \textit{boson stars}~\cite{Jetzer:1991jr}; the global $U(1)$ charge guarantees their stability. In the regime in which gravity is negligible the same system leads to non-topological solitons such as \textit{Q-balls}~\cite{Coleman:1985ki}: they can be formed if the attractive self-interactions of the complex scalar field compensate for the gradient energy. Contrary to fermion stars (like neutron stars) for which the mass of the compact object is $M_f\sim M_p^3/m^2$ (where $M_p$ is the reduced Planck mass and $m$ the mass of the corresponding particle), typical boson stars have a much smaller mass $M_b\sim M_p^2/m$ and are therefore sometimes called {\it mini-boson stars}~\cite{Jetzer:1991jr}. If the self-interactions are strong enough, i.e. if $g M_p^2/m^2 \gg 1$ (the coupling $g$ is defined in the scalar potential in footnote~\ref{foo:Potential}), the star can be as heavy as the corresponding fermion star and then it is simply called boson star~\cite{Colpi:1986ye}.

In the case of real scalar fields, there is no global symmetry to guarantee the stability of the compact objects, but these can be long-lived due to approximate symmetries~\cite{Mukaida:2016hwd}. The object in question in this case is a \textit{pseudo-soliton}, whose main representative is given by an \textit{oscillaton}. Typical examples of oscillatons are \textit{axion stars}\footnote{Axion stars are expected to be produced in the unbroken PQ scenario~\cite{Kolb:1993zz, Kolb:1993hw, Hogan:1988mp, Visinelli:2017ooc} and for this reason they are particularly well motivated.}. \textit{Oscillons}~\cite{Seidel:1991zh, Kolb:1993hw, Alcubierre:2003sx, UrenaLopez:2001tw, UrenaLopez:2002gx, UrenaLopez:2012zz} belong to the same class as oscillatons, but they are typically restricted to the regime in which the role of gravity is negligible. Their stability is provided by attractive non-linear interactions, rather than gravity: their existence requires the scalar potential to be \textit{shallower than quadratic} at least on one side around the minimum. Oscillons are formed during a preheating-like stage in a wide variety of inflationary models~\cite{Amin:2011hj, Amin:2014eta, Amin:2013ika, Amin:2010jq, Antusch:2016con, Antusch:2017flz}, as well as in string moduli potentials as a consequence of field displacement~\cite{Antusch:2017flz}. We collectively refer to pseudo-solitonic objects arising from a single real scalar field, i.e. the bottom line of Tab.\ \ref{tab:Classification}, as \textit{real scalar stars} (RSSs). The goal in this paper is to explore the importance of gravity for the stability of RSSs, studying their dynamics at the border between the oscillon and oscillaton regimes.

The fate of RSSs is the subject of the analysis in the present paper. It is known that in the free-field case the solutions are characterized by a stable and an unstable branch~\cite{UrenaLopez:2002gx}: perturbed configurations belonging to the latter either collapse to black holes or they migrate to the stable branch, depending on the sign of the perturbation. In the interacting case, studying the equilibrium configurations is challenging both from the analytic and from the numerical points of view~\cite{UrenaLopez:2012zz}. It is, however, possible to address this question performing numerical studies that include the effects of General Relativity (GR) and the full non-linear dynamics of the self-interactions\footnote{See~\cite{Cotner:2018vug} for an alternative approach.}.

An example of such a study is~\cite{Helfer:2016ljl}, where the authors considered stability of RSSs in the specific case of an axion potential. As expected they found that, depending on the parameters of the model, axions stars can be stable, disperse or collapse to black holes. Interestingly, non-linear interactions are crucial for a proper understanding of the dynamics of these objects: even if the initial radius of the star is much larger than the corresponding Schwarzschild radius, the interplay of non-linear interactions and gravity can drive the star to collapse or to dispersion. 

\begin{table}[h!]
\centering
\begin{tabular}{|c|c|c|c|}
\hline
\textbf{Scalar} & {\bf $G = 0$} &  \multicolumn{2}{c|}{\bf $G = 1$} \\ 
\hline \hline
\multirow{2}{4em}{\centering{Complex}}  & \multirow{2}{6em}{\centering{\textbf{\textit{Q-Balls}}} \\ \centering{Global $U(1)$}} & \textbf{\textit{Mini-Boson Stars}} & \textbf{\textit{Boson Stars}} \\
 &  & weak self-interactions & strong self-interactions \\
\hline\hline
\multirow{2}{2em}{Real} & \textbf{\textit{Oscillons}} &  \multicolumn{2}{c|}{\multirow{2}{6em}{\textbf{\textit{Oscillatons}}}} \\ 
& attractive self-interactions & \multicolumn{2}{c|}{} \\  
\hline
\end{tabular}
\caption{Classification of complex and real scalar stars. Here and in the rest of the paper we denote by ``$G = 0$" the cases in which gravity effects are negligible (where $G$ is the Newton's constant), and by ``$G = 1$" the cases in which gravity effects are important.}
\label{tab:Classification}
\end{table}

RSSs formed after inflation and before BBN and composed of the inflaton or any other modulus that is displaced from the late-time minimum after inflation~\cite{Coughlan:1983ci, deCarlos:1993wie, Banks:1993en, Kane:2015jia} are particularly well-motivated because - when gravity is negligible - the existence of these compact objects requires the potential to be shallower than quadratic, which translates into an attractive interaction between particles. This requirement is met both by inflationary potentials, as the latest Planck 2018~\cite{Akrami:2018odb} results favour plateau-like potentials, and by moduli potentials, since string models feature potentials that open-up on at least one side with respect to the late-time minimum. As we will show in the rest of the paper, the most interesting dynamics appear	 if the effects of gravity become of the same order as those due to self-interactions.

Note that the objects we consider in this work have a compactness\footnote{The compactness is defined as $\mathcal{C} = M/R$, where $M$ is the total mass of the RSS and $R$ its radius, containing $90\%$ of the mass of the star.} comparable to that of the corresponding black hole. The formation of such objects needs to be checked for each specific model via dedicated lattice simulations. In the simplest and most model-independent scenarios, self-interactions of a single field are sufficient to make the quantum fluctuations grow and enter the non-linear regime. The growth can take place mainly through \textit{parametric resonance} (see~\cite{Amin:2014eta} and references therein) or \textit{tachyonic resonance}~\cite{Brax:2010ai}. Focusing on the models studied in the present paper, parametric and tachyonic resonance are efficient production mechanisms in the case of the KKLT model and $\alpha$-attractors T-models respectively\footnote{For $\alpha$-attractor T-models~\cite{Kallosh:2013yoa, Kallosh:2015lwa} the region of the parameter space investigated is borderline in terms of the efficiency of the tachyonic resonance production mechanism, and should be checked through lattice simulations. The formation of RSSs is however beyond the scope of this work and we leave it for the future.}. In the case of $\alpha$-attractor E-models (or Starobinsky-like potentials), the production might be difficult to achieve through parametric and/or tachyonic resonance, but it could take place through other mechanisms, e.g. \textit{i)} through parametric resonance induced by a second oscillating field~\cite{Kofman:1997yn}, \textit{ii)} through some enhancement in the scalar power spectrum, as for the formation of primordial black holes (see~\cite{Widdicombe:2018oeo} for recent work in this direction), \textit{iii)} through dynamical clustering of lighter objects~\cite{Amin:2019ums}.

A better understanding of the evolution of moduli stars is important for a number of reasons:
\begin{enumerate}
\item Non-spherically symmetric oscillons produce GWs due to the dynamics of each single object~\cite{Dufaux:2007pt}. The GW spectrum at production is peaked at frequencies $f \sim \mathcal{O}(m)$, where $m$ is the mass of the scalar field. The current diluted value is roughly given by
\begin{equation}
f \gtrsim \left(\frac{m}{\text{TeV}}\right)^{5/6} \, \text{Hz} \,,
\end{equation}
where the uncertainty is related to the knowledge of the exact production time and is removed by numerical simulations. For $m \sim 10^9 \, \text{GeV}$ the peak would fall in the LIGO frequency range. In~\cite{Antusch:2017vga}, the authors parameterized the oscillon profile as
\begin{equation}
\label{eq:ParametrizedProfile}
\phi(t, \mathbf{x}) = \Phi(t) \mathcal{F}(t, \mathbf{x}) \,,
\end{equation}
where $\Phi(t) \equiv \mathcal{A} f(t)$ is the time-dependent oscillon amplitude ($f(t)$ a periodic function, e.g. $f(t) = \cos(t)$), while $\mathcal{F}(t,\mathbf{x})$ is the asymmetric profile. Since the source of the anisotropic stress tensor is $\propto \partial_i \phi \partial_j \phi$, the energy density in GWs produced by such a configuration is
\begin{equation}
\label{eq:GWSpectrumAmplitude}
\Omega_{\text{GW}} \propto \mathcal{A}^4 \,,
\end{equation}
hence the amplitude of the GW spectrum is proportional to the fourth power of the oscillon field amplitude. Typically, the larger is the oscillaton field amplitude, the larger is the compactness $\mathcal{C}$ of the compact object. In this paper we focus on the dynamical evolution of spherically symmetric configurations, and we leave the analysis of the formation of dense RSSs through various mechanisms operating in the early Universe and of the related GW production to future work.
\item If RSSs are stable for a long time, they could undergo dynamical interactions including the \textit{i)} formation of binaries~\cite{Amin:2019ums} and collisions~\cite{Helfer:2018vtq} with subsequent emission of primordial GWs that would result in a stochastic background; \textit{ii)} clustering and formation of heavier bound objects~\cite{Belotsky:2018wph}, with possible collapse to heavy black holes that could survive even after the decay of the fields that compose the original star, and constitute a dark matter candidate. The richness of these dynamics would be enhanced if the stars collapse to black holes, since the longer lifetime would make it easier for the black holes to cluster. RSSs themselves could constitute a dark matter candidate if the mass of the scalar field is so small that they have not decayed yet, see~\cite{Olle:2019kbo} for recent work in that direction. In the regime of weak gravity, the RSS would have a low density and a very large size, composing the core of dark matter halos and possibly providing a solution to the cusp-core problem of cold dark matter, see e.g.~\cite{Marsh:2015wka}.
\item If RSSs collapse to black holes, these would have a very small mass and would evaporate quickly, providing one of the necessary conditions required for an explicit realization of the \textit{Hawking genesis} scenario proposed in~\cite{Lennon:2017tqq}. As black holes evaporate democratically into all sectors of the model, such a scenario would provide strong constraints on string model building, given that hidden sectors typically contain dark radiation candidates~\cite{Cicoli:2012aq, Cicoli:2015bpq, Acharya:2015zfk, Halverson:2018xge, Cicoli:2018cgu}, which is highly constrained by CMB measurements~\cite{Aghanim:2018eyx}. Moreover, the collapse to black holes would modify substantially the lifetime of the compact objects. While RSSs have a lifetime which is $\tau_{\text{star}} \sim 10^3-10^4 \times m^{-1}$, the lifetime for a black hole with mass\footnote{We take the mass of the black hole to be of the same order of the mass of the original star, which is consistent with our findings in collapsing cases.} $M \sim M_p^2/m$ would be
\begin{equation}
\tau_{\text{BH}} \sim \frac{M^3}{M_p^4} \simeq \left(\frac{M_p}{m}\right)^2 \times m^{-1} \,,
\end{equation}
which is larger than $\tau_{\text{star}}$ for any $M_p/m \gtrsim 10^{2}$.
\end{enumerate}

A genuinely new feature of some of the models that we consider in the present paper is the asymmetry of the scalar potential around the minimum. While in symmetric cases the real scalar particles feel either an attractive (if the potential is shallower than quadratic) or a repulsive (if the potential is steeper than quadratic) force, in the case of asymmetric potentials the two behaviours alternate at each oscillation. We would have naively expected this feature to make the RSS more unstable if compared to the symmetric case. We will show in the next subsection that this is actually a model-dependent statement, emphasizing the importance of performing numerical simulations to find out the true RSS dynamics when non-linear interactions and gravity are important.

The paper is organized as follows: in Sec.~\ref{sec:Models} we review the models for which we will analyze the dynamics of the RSSs. In Sec.~\ref{sec:NumericalSetup} we briefly summarize the numerical setup that we use for the simulations in the relativistic regime. In Sec.~\ref{sec:Results} we illustrate the results of the simulations. In Sec.~\ref{sec:Conclusions} we draw conclusions and make suggestions for future work.

\section{Models}
\label{sec:Models}

In this Section we introduce the models for which the dynamics will be studied.
In order to determine whether a given potential supports these solutions in the first place let us  consider a single real scalar field $\phi$ in the absence of self-gravity for the non-perturbative scalar field clumps. One crucial parameter is the typical scale of the potential $\Lambda$, which sets the size of the non-linear interactions. Take the for instance the potential
\begin{align}
\label{eq:GenericPotential}
V &= \frac{1}{2} m^2 \phi^2 + \frac{1}{3!} \lambda \phi^3 + \frac{1}{4!} g \phi^4 \equiv \nonumber  \\
&\equiv m^2 \Lambda^2\left[ \frac{1}{2} \left(\frac{\phi}{\Lambda}\right)^2 + \frac{1}{3!} \frac{\lambda \Lambda}{m^2} \left(\frac{\phi}{\Lambda}\right)^3 + \frac{1}{4!} \frac{g \Lambda^2}{m^2} \left(\frac{\phi}{\Lambda}\right)^4 \right] \,.
\end{align}
We define the typical scale of the potential $\Lambda$ as the scale at which the potential deviates significantly from a quadratic potential. Requiring that the interaction terms are of the same order of the mass term in the potential at $\phi \sim \Lambda$, this corresponds to requiring that the couplings are of order
\begin{equation}
\lambda \sim m \left(\frac{m}{\Lambda}\right) \,, \qquad g \sim \left(\frac{m}{\Lambda}\right)^2 \,.
\end{equation}
The potential in Eq. \eqref{eq:GenericPotential} supports RSS (oscillon, i.e. solutions in which gravity is negligible) if~\cite{Hertzberg:2010yz}
\begin{equation}
 \frac{5}{3} \left(\frac{\lambda}{m}\right)^2 - g > 0\ .
 \label{eq:formCond}
 \end{equation}
As a rule of thumb, if the potential is symmetric around the minimum ($\lambda = 0$), it supports oscillon solutions whenever it is ``shallower than quadratic", which translates into attractive interactions. In order to estimate whether or not self-gravity is negligible for these solutions, let us restrict ourselves for a moment to the simpler case $\lambda = 0$: the mass and radius of the RSS in the dense regime, i.e. when the oscillation amplitude of the field is $\phi \sim \mathcal{O}(\Lambda)$, take the form~\cite{Hertzberg:2010yz}
\begin{equation}
M \simeq \frac{m}{g} \simeq \frac{\Lambda^2}{m} \,, \qquad R \simeq \frac{1}{m} \,.
\end{equation}
In order to have a measure of the importance of gravity for the problem at hand, we can take the ratio between the RSSs radius $R$ and the corresponding Schwarzschild radius $R_s$:
\begin{equation}
\frac{R_s}{R} \approx \left(\frac{\Lambda}{M_p}\right)^2 \,,
\end{equation}
which suggests that GR effects become non-negligible if the typical scale of the potential $\Lambda$ is close to $M_p$. It follows immediately that for RSSs formed in the \Kahler moduli inflation scenario~\cite{Conlon:2005jm}, as shown in~\cite{Antusch:2017flz}, GR effects are never important, since the typical scale of the potential is set by the string scale $\Lambda \equiv M_s = M_p/\mathcal{V}^{1/2}$, where $\mathcal{V} \gg 1$ is the volume of the extra-dimensions.\\
Note that the criterion described above is relevant for the need to include GR effects as we do in this paper - objects with compactness $\mathcal{O}(1)$ (and down to around $\mathcal{O}(10^{-3}))$ can show relativistic behaviour during their evolution. But gravity can already provide support for RSSs at much lower compactnesses, and then Newtonian gravity is already necessary (but also sufficient) for their study, see e.g.~\cite{Visinelli:2017ooc, Schiappacasse:2017ham}.

In this paper we study various examples of potentials, with highly compact RSSs, in the regime in which GR effects are non-negligible. As already mentioned, we do not deal with the RSS formation mechanism, and we leave the analysis of the effects of GR on such mechanisms to future work. As shown in~\cite{Amin:2019ums}, it is likely that throughout most of the formation and evolution, Newtonian gravity would be sufficient for a study of the evolution from oscillon to oscillaton. It is only in extreme regimes, or in cases in which perturbations are enhanced, that the objects become sufficiently compact for GR effects to be important. One can then ask the question, at what point do such objects become unstable to gravitational collapse? This question is non trivial, especially in cases in which the potential is asymmetric about the minimum, where the object will experience alternating dispersive and attractive interactions during the course of one oscillation. We thus focus on the stability of large perturbations in the field, and the interplay between the non-linear field dynamics, and gravitational effects\footnote{For a similar study in the context of complex field boson stars see~\cite{Cunha:2017wao}.}. 

\begin{figure}[htp!]
    \centering
    \includegraphics[width=1.00\textwidth]{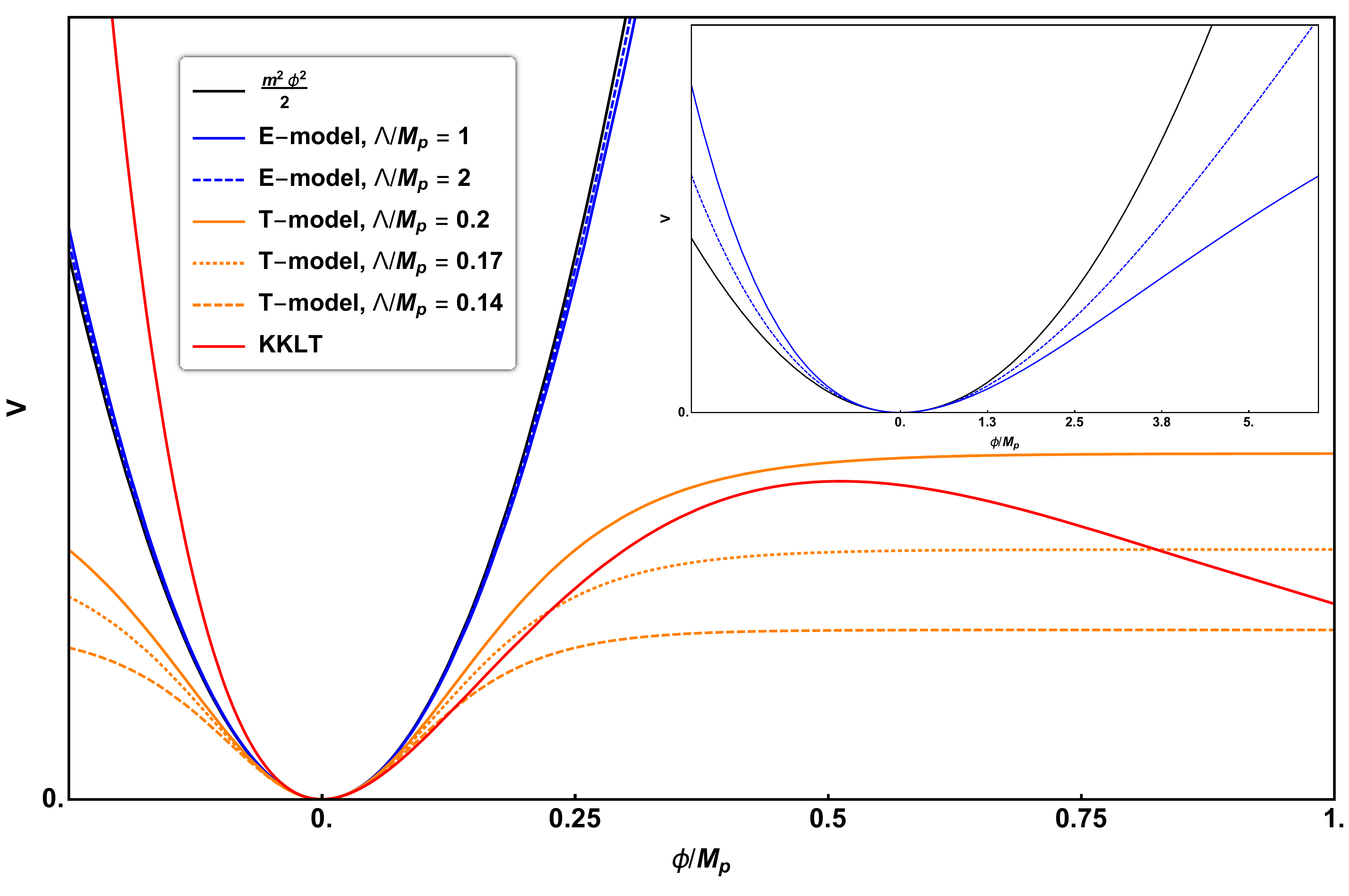}
    \caption{Summary of the potentials analyzed in Sec.~\ref{sec:Results}.}
\label{fig:Potentials}
\end{figure}

We will consider three classes of potentials: $\alpha$-attractor T- and E-models, and KKLT. The shapes of the potentials, crucial for the dynamics of the RSSs, is summarized in Fig.~\ref{fig:Potentials} where we normalized the mass of the field to the same mass $m$. Although the results illustrated in Sec. \ref{sec:Results} are valid for any value of the mass $m$, for $\alpha$-attractor potentials, once $\Lambda$ is fixed, then also the value of the mass $m$ is fixed in order for the inflationary dynamics to be in agreement with observations.\\

\noindent \textbf{$\alpha$-attractor T-models}\\
\noindent $\alpha$-attractors have been proposed as supergravity models of inflation~\cite{Kallosh:2013yoa, Kallosh:2015lwa}. In this class of models the field-space curvature, parametrised by $\alpha$, plays a crucial role in the determination of the inflationary observables, as it allows for the tuning of the tensor to scalar ratio. The general scalar potential for $\alpha$-attractor T-models takes the form
\begin{equation}
V = V_0 \, \tanh^{2n} \left(\frac{\phi}{\Lambda}\right) \,,
\end{equation}
where $V_0$ is a normalization constant and $\Lambda$ is the typical scale of the potential, related to the geometrical parameter $\alpha$ via $\frac{\Lambda}{M_p}\propto \sqrt{\alpha}$. Given that the tensor-to-scalar ratio can be written as 
\begin{equation}
r\approx \frac{2}{N_e^2} \frac{\Lambda^2}{M_p^2} \,,
\end{equation}
where $N_e\equiv \ln a$ is the number of efoldings of expansion during inflation, the current upper bound $r\le 0.07$ constrains the scale $\Lambda$ to $\Lambda/M_p < 10$ for $N_e\sim 50-60$. This corresponds to having the observable range of inflationary expansion being produced in the plateau region of the potential.

Normalisation of the scalar curvature perturbations 
\begin{equation}
\mathcal{A}_s=\frac{H^2}{8 \pi^2 \epsilon}\sim 2\times 10^{-9}
\label{eq:As}
\end{equation}
implies that the scale of the potential $V_0$ is bounded from above: $V_0\le 10^{-9} M_p^4$.

Around the minimum the potential admits an expansion of the form
\begin{equation}
V=V_0 \left(\frac{\phi}{\Lambda}\right)^{2n} \left\{1-\frac{2n}{3}\frac{\phi^2}{\Lambda^2}+\frac{7 n+10 n^2}{45 }\frac{\phi^4}{\Lambda^4}+\mathcal{O}(\phi^6)\right\}
\end{equation}
and therefore it obeys the ``shallower than quadratic" criterion for the existence of oscillon solutions in the absence of gravity. We stress that this criterium is simply an indication for when the formation might take place in the absence of gravity. If gravity effects are included meta-stable solutions exist even for the free-field case. In our simulations we use the full potential for all models. We will set $n=1$, in which case the cubic coupling vanishes and the quartic becomes $g = - 16 V_0/\Lambda^4$, automatically satisfying Eq. \eqref{eq:formCond}. In this kind of potential oscillons are formed for $\Lambda \ll M_p$ in the broad resonance regime \cite{Amin:2019ums}.\\ 

\noindent \textbf{$\alpha$-attractor E-models}\\
\noindent $\alpha$-attractor E-models of inflation feature a plateau-like potential which is favoured by the latest Planck results~\cite{Akrami:2018odb}. They are a generalization of the Starobinsky potential of the form
\begin{equation}
\label{eq:StarobinskyPotential}
V = V_0 \left(1 - e^{-\frac{\phi}{\Lambda}}\right)^2 \,.
\end{equation}
Notice that the original Starobinsky model has $\Lambda/M_p = \sqrt{3/2}$ while {fibre inflation} corresponds to the case with $\Lambda/M_p = 1/\sqrt{3}$ \cite{Cicoli:2008gp}. Like the T-models, this class of models features a minimum at the origin and a flat plateau for $\phi\gg \Lambda$. We note however that while the minimum for the T-models is symmetric, in the case of E-models it is not.

If  $\phi$ is the inflaton field,  to leading order in $N_e\gg1$ one has
\begin{equation}
r\approx\frac{8}{N_e^2} \left(\frac{\Lambda}{M_p}\right)^2\qquad\text{and}\qquad \mathcal{A}_s\approx \frac{N_e^2}{12 \pi^2} \frac{V_0}{\Lambda^2 M_p^2}\ .
\end{equation}
Inflationary constraints on $r$ and $\mathcal{A}_s$ map to the bounds $\Lambda/M_p \leq 5$ and $V_0\le 10^{-9} \, M_p^4$. We stress that these only apply if $\phi$ is responsible for inflation and that in Sec.~\ref{sec:Results} we will consider cases where these conditions are violated.

Around the minimum one may expand
\begin{equation}
\frac{V}{V_0}=\left(\frac{\phi}{\Lambda}\right)^2-\left(\frac{\phi}{\Lambda}\right)^3+\frac{7}{12}\left(\frac{\phi}{\Lambda}\right)^4+\mathcal{O}\left(\left(\frac{\phi}{\Lambda}\right)^6\right)
\end{equation}
and so the E-model potential is shallower than quadratic to the right of the minimum and steeper to its left, c.f. Fig. \ref{fig:Potentials}. As a consequence of this asymmetry, over the course of a cycle, the RSS particles will experience alternating attractive and repulsive forces.

This class of potential therefore meets the zeroth order criterion for oscillon formation in the absence of gravity. If the oscillon amplitude is small enough such that the quartic expansion of $V$ is a good approximation, the  formation criterion of  Eq. \eqref{eq:formCond} reduces to $V_0 >7/30 \, \Lambda^2 M_p^2$. Note that in this regime $\phi$ cannot be the inflaton as it would violate the normalisation of the scalar power spectrum, Eq. \eqref{eq:As}.\\

\noindent \textbf{KKLT}\\
\noindent KKLT is one of the most well-known mechanisms for K\"ahler moduli stabilisation in Type IIB string compatifications~\cite{Kachru:2003aw}. In its  simplest variant it features one single \Kahler modulus $T = \tau + i a$ describing the overall size of the compactification space. The scalar potential for the complex scalar $T $ is generated by non-perturbative corrections to the action. More explicitly it is generated by the \Kahler potential and superpotential 
\begin{equation}
\label{eq:KahlerSuperPotential}
K \supset -3 \ln\left(T + \overline{T}\right) \,, \qquad W = W_0 + A e^{-a_N T} \,,
\end{equation}
where $W_0$ is the Gukov-Vafa-Witten superpotential arising from fluxes~\cite{Gukov:1999ya}, $A$ is a $\mathcal{O}(1)$ coefficient depending on the details of the non-perturbative physics that produces the non-perturbative correction and $a_N = \frac{2 \pi}{N}$, where $N$ is the number of D7-branes wrapping the 4-cycle whose volume is parameterized by $\tau$. 
Given that Eq. \eqref{eq:KahlerSuperPotential} implies
\begin{equation}
\mathcal{L}\supset \frac{3}{4} \left(\frac{\partial\tau}{\tau}\right)^2=\frac{1}{2} (\partial \phi)^2,
\end{equation}
the canonically normalised volume modulus is
\begin{equation}
\frac{\phi}{M_p}=\sqrt{\frac{3}{2}}\ln\tau .
\label{eq:cannorm}
\end{equation}
Assuming that the axion $a$ is stabilised at the minimum of its scalar potential, the potential for $\tau$ arising from Eq. \eqref{eq:KahlerSuperPotential} takes the form
\begin{equation}
V = \frac{V_0}{6 \tau^2} \left[a A^2 (3 + a \tau) e^{-2 a \tau} - 3 a A W_0 e^{-a \tau}\right] + V_{\rm dS} \,,
\end{equation}
where $V_{\rm dS} = V_{\rm up} \, \tau^{-2}$ is an additional term that ensures a de Sitter minimum~\cite{Kachru:2003aw}. This potential features a minimium with vanishing cosmological constant at $\bar{\tau}$ defined by 
\begin{equation}
W_0=-\frac{1}{3}A e^{-a \bar{\tau}}(5-2 a \bar{\tau}).
\end{equation}
Around the minimum the potential admits an expansion of the form
\begin{equation}
\begin{split}
V=&\frac{a^3 A^2 e^{-2 a \bar{\tau}}}{12 \bar{\tau}^2}\left\{(3+2 a \bar{\tau})(\tau-\bar{\tau})^2+\left(\frac{-19 }{3}a-\frac{6}{\bar{\tau}}-2 a^2 \bar{\tau}\right)(\tau-\bar{\tau})^3\right.\\
&\left.+\left(\frac{59}{12}a^2+\frac{9}{\bar{\tau}^2}+\frac{32 a}{3 \bar{\tau}}+\frac{7 a^3 \bar{\tau}}{6}\right)(\tau-\bar{\tau})^4+....\right\}\ .
\end{split}
\label{eq:VKKLMTmin}
\end{equation}
From Eq. \eqref{eq:VKKLMTmin} and Eq. \eqref{eq:cannorm} one sees that $V$ is asymmetric around the minimum and shallower than quadratic to the right of it, meeting the zeroth order criterium for the existence of oscillon solutions in the absence of gravity. It has been shown in~\cite{Antusch:2017flz} that in the large $|W_0|$ regime\footnote{The minimum in KKLT is located at $|W_0| \sim e^{-a \tau} \ll 1$. The adjective \textit{large} here means $10^{-6} \lesssim |W_0| \lesssim 10^{-2}$.} oscillons are formed in the KKLT scenario.

\section{Numerical Setup}
\label{sec:NumericalSetup}

We use numerical relativity (NR) simulations to search for stable solutions to the Einstein-Klein-Gordon (EKG) equations with each of the given models. We start with profiles for the scalar field which would, in the absence of self-interactions in the field, provide stable solutions. For each of the models described above, we evolve these forward in time to see if a stable state is reached, or whether collapse or dispersion of the perturbation occurs.

In this Section we give further details of the initial conditions and numerical techniques used, including code verification tests.

\subsection{Initial conditions}

In the free-field oscillaton case, with $V(\phi) = \frac{1}{2} m^2 \phi^2$, meta-stable solutions\footnote{As emphasized in Sec.~\ref{sec:Introduction}, these compact objects are bound to disperse, given that there is no conservation law preventing their decay, see however~\cite{Gleiser:1999tj, Salmi:2012ta, Gleiser:2008ty, Saffin:2006yk, Gleiser:2019rvw}.} to the EKG equations can be found numerically using a shooting method, as in~\cite{Alcubierre:2003sx,UrenaLopez:2002gx,UrenaLopez:2001tw}. Once the field mass $m$ is fixed, they describe a one parameter family of solutions, parameterised by the central amplitude or, equivalently, by the total mass of the perturbation. We will use the latter to specify the initial configurations of the simulations. The one parameter family of solutions is composed of two different branches, i.e. the \textit{stable branch} and the \textit{unstable branch}. For a fixed value of $m$, upon increasing the central amplitude of the field, the total mass $M$ of the equilibrium configurations grows till the maximum value $M_{\rm max} = 3.04 \, M_p^2/m$: this is the stable branch. If the central amplitude is increased any further, the total mass $M$ starts moving towards smaller values: this is the unstable branch. Perturbed configurations on the stable branch quickly go back to a stable configuration, while perturbed configurations on the unstable branch either collapse to black holes (if the perturbation increases the total mass for fixed central amplitude) or it migrates towards the stable branch (if the perturbation decreases the total mass for fixed central amplitude). We will consider configurations that belong to the stable branch in the free-field case, in the range of total masses $2.6 \, M_p^2/m < M < 3.0 \,M_p^2/m$. We choose this range based on the following observations: \textit{i)} below $2.6 \, M_p^2/m$ GR effects become less and less important for the potentials considered in this paper, and \textit{ii)} above the mass $M = 3.04 \, M_p^2/m$ any perturbation makes the free-field RSS unstable, so this value sets our upper bound for the total mass. For definiteness, we consider the three values $M = 2.6 \, M_p^2/m$, $M = 2.9 \, M_p^2/m$ and $M = 3.0 \, M_p^2/m$ for the initial total mass of the RSSs, respectively corresponding to the initial central amplitudes $\phi/M_p = 0.2$, $\phi/M_p = 0.3$, $\phi/M_p = 0.4$ for the scalar field of the free-field RSSs. In this way, we quantify and compare the change in stability as a result of the non-linear self-interactions.

A time slice of these solutions is used as the initial conditions for our simulations. For numerical convenience, we choose a slice on which $\phi(r) = 0$ and $\dot \phi(r) \neq 0$, see App. A of~\cite{Helfer:2016ljl} for further details. This gives us a moment of time symmetry (at the level of the metric and the stress-energy tensor) such that the momentum constraint is trivially satisfied. In addition, since the field is everywhere at the minimum in the potential, it does not at this point ``feel'' the deviations from the free-field case, and thus the numerically obtained RSS solutions for the metric and the field satisfy the Hamiltonian constraint exactly, without further modification.

Whilst it should be emphasised that the free-field oscillaton solutions are not necessarily stable once self-interactions are added, they nevertheless provide a convenient set of initial conditions with which to systematically test the stability properties of each model. The interacting solution is likely to be similar to the non-interacting one, at least for small self-interactions, so one may hope that such an approach makes it more likely that we will converge on the stable solutions (compared to starting with, for example, a gaussian, or other random perturbation profiles\footnote{Although note that in \cite{Michel:2018nzt}, gaussians were used successfully to investigate stable and unstable solutions, and yielded results similar to those found in \cite{Helfer:2016ljl}, which used a similar approach to our own.}), if such solutions exist. We note that a failure to settle into a steady state in our simulations is not evidence that no stable solution exists - we could simply have not found it. However, in all cases studied, we do see our solution settle into a stable state below some critical mass, justifying the approach.

A further practical advantage of using the RSS solutions is that one can check that in the limit of small amplitudes and/or small self-interactions, the solution remains stable and unchanged (modulo some initial gauge evolution), which is an invaluable check of the code and potential function.

Finally we note that, for asymmetric potentials, we have a choice of which direction in field space we initially send the field - to the attractive or repulsive part of the potential. We do not find that this makes a significant difference to our results; the collapsing cases will still collapse, just with a longer or shorter timescale. So we choose to always send it in the direction of positive $\phi$.

\subsection{Numerical methods}

We use the publicly available NR code $\textsc{GRChombo}$ to evolve the initial data forward in time. GRChombo is itself built on top of the open source $\mathtt{Chombo}$ framework \cite{Chombo}. For a more full discussion of $\textsc{GRChombo}$ see \cite{Clough:2015sqa}, or the website \url{www.grchombo.org}. For completeness we describe below the key features of the code as used in this work.

The evolution uses the method of lines, with metric and field derivatives calculated using finite difference stencils of the grid values and a Runge-Kutta time integration for the evolution equations. The metric in the 3+1D decomposition:
\begin{equation}
ds^2 = - \alpha^2 dt^2 + \gamma_{ij} (dx^i + \beta^i dt) (dx^j + \beta^j dt) \,,
\end{equation}
where $\alpha$ is the lapse and $\gamma_{ij}$ is the three-metric on the equal time hypersurfaces.

The BSSN formulation \cite{Nakamura:1987zz,Shibata:1995we,Baumgarte:1998te} of the 3+1D ADM equations \cite{Arnowitt:1962hi} is used for numerical stability, with CCZ4 \cite{Alic:2011gg} constraint damping terms added to reduce constraint violation over the simulation. The moving puncture gauge  \cite{Pretorius:2005gq,Baker:2005vv,Campanelli:2005dd} is employed to allow the simulations to follow black hole formation - this is a dynamical gauge choice which maintains coordinate observers at approximately fixed positions relative to the center of the domain, avoiding the focussing of geodesics on the overdense regions.

The length of the domain is $L = 64 \times 1/m$, where $m$ is the mass of the field, and we enforce between 4 and 7 (2:1) refinement levels at the start of the simulation with the coarsest having $N$ grid points, where $N=64^3$ is usually sufficient, but several values are used to check convergence. Additional resolution is added if collapse occurs, triggered dynamically by second derivatives in the field $\phi$ and metric conformal factor $\chi$ exceeding set thresholds. The metric conformal factor is defined as $\chi = \left(\det \gamma_{ij}\right)^{-1/6}$, thus the formation of a singularity is signalled by $\chi$ falling to zero at the center, where the metric determinant becomes infinite. Note, however, that due to the rapidly varying amplitude of the field, we frequently needed to fix the resolution at the maximum level achieved using Adaptive Mesh Refinement (AMR), following an initial exploratory run, so as to control the errors from too frequent regridding.
Kreiss-Oliger dissipation is used to control numerical errors in the evolution of the fields, in particular that arising at the grid boundaries. We evolve only 1/8th of the grid using the octant symmetry in cartesian coordinates, and apply Sommerfeld boundary conditions (which permit outgoing radiation and thus reduce reflections into the grid) at the outer edges. Whilst a code adapted to the spherical symmetry of the solutions would provide a far more efficient approach, the method adopted gives us the potential to generalise our studies in future work. Beyond tracking the dynamics of $\chi$, we check the formation of black holes by finding the apparent horizon on each spatial hypersurface \cite{Thornburg:2006zb}. Apparent horizons are ``trapped" surfaces for which the expansion of the outgoing null geodesics normal to the surface is zero - for a stationary metric they coincide with the event horizon but in a dynamical spacetime the event horizon (which can only be found by integrating the null geodesics over all time) may lie outside the apparent horizon. An apparent horizon area thus provides a lower limit on the mass of any black hole present in the spacetime.

\begin{figure}
    \centering
    \includegraphics[width=0.49\textwidth]{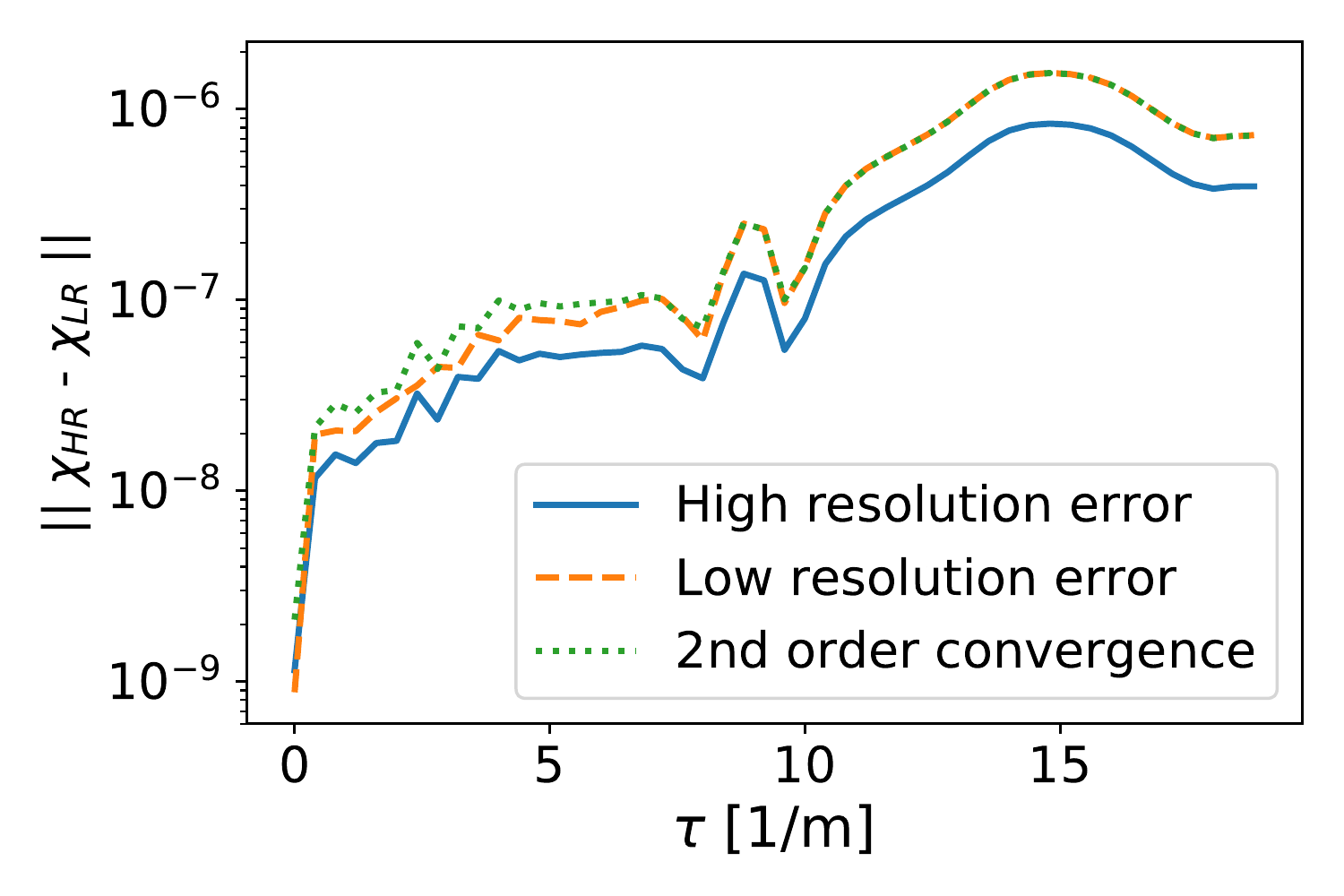}
    \includegraphics[width=0.49\textwidth]{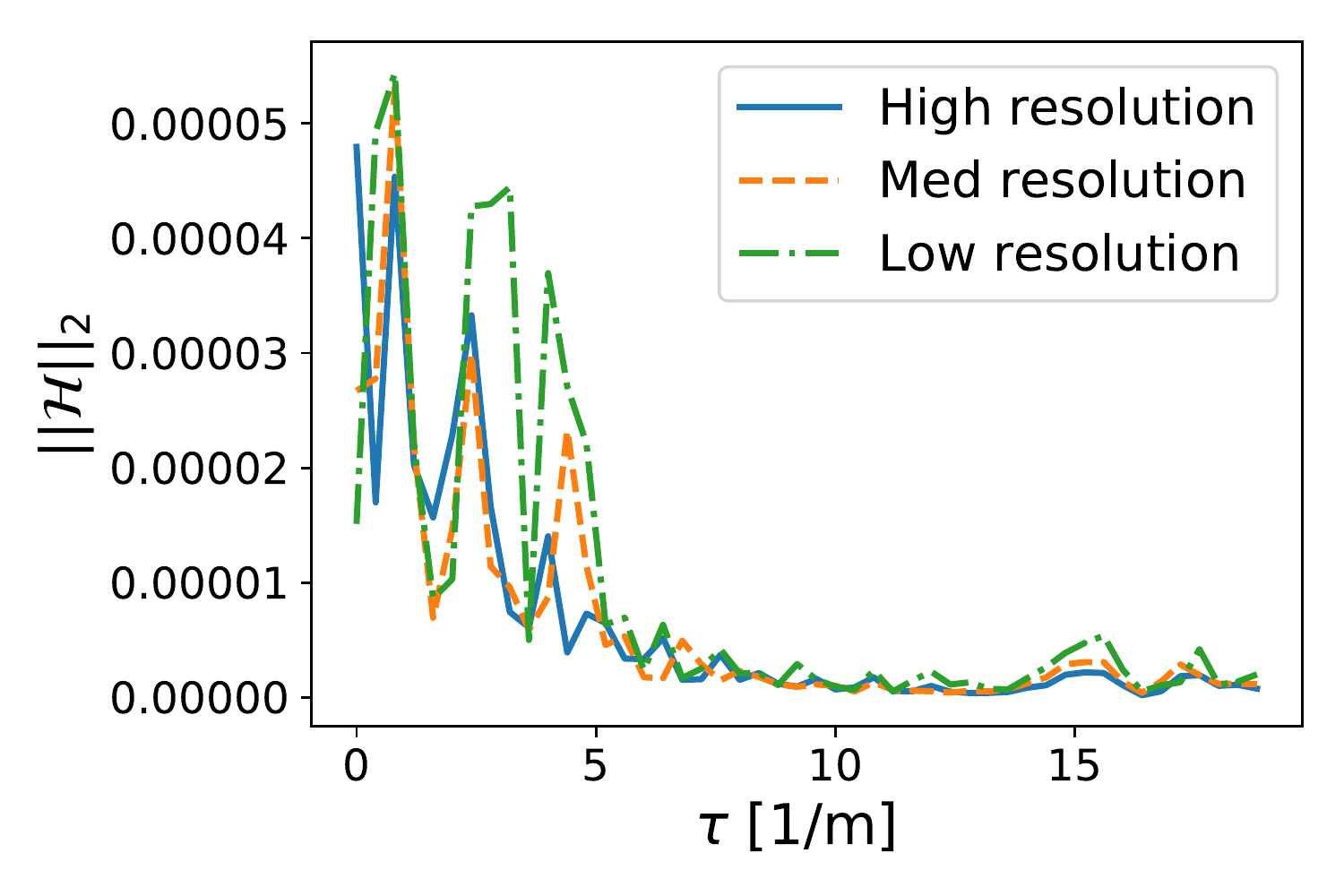}
    \caption{Here we show two plots from our convergence testing for the T-model potential ($n=1$) with $\Lambda/M_p = 0.2$ and total mass of the initial configuration $M = 2.9 \, M_p^2/m$. The first shows that the error in the conformal factor $\chi$ is converging at approximately 2nd order, calculated by comparing the difference between the results at 3 successive resolutions, which correspond to base resolutions of $64^3$, $80^3$ and $96^3$, each with 7 levels of 2:1 refinement. On a plot of $\chi$ versus time the results at the three resolutions are indistinguishable. The second plot shows the L2 norm of the Hamiltonian constraint violation at each resolution, showing that it remains bounded thoughout the simulation and decreases with resolution.}
\label{fig:Converge}
\end{figure}

We have checked convergence of the simulations in critical cases where we identify a boundary between collapse and stability. An illustration of the tests performed is shown in Fig. \ref{fig:Converge}. We find the solution to be well converged in the range of base resolutions between $N = 48^3$ to $N = 96^3$. In other simulations away from the critical cases we check that the profiles obtained for the field $\phi$ and the inverse determinant of the metric $\chi$ are indistinguishable when the simulation is performed at several different base resolutions centered on $N = 64^3$.

We note that the time period of our simulations is necessarily much shorter than the expected lifetime of the oscillatons due to computational limitations, such that we cannot explicitly confirm that the solutions we class as ``meta-stable'' will remain so in the longer term. However, where we see the solution settle into a periodic pattern (usually with modulations that appear to be decaying over time), it is reasonable to assume that in the absence of further perturbations it will then slowly decay rather than suffer a sudden collapse. Running several simulations for longer periods supports this intuition.
 
Finally, we also present cases with gravity ``switched off" for comparison. In such cases we evolve only the scalar field dynamics on a Minkowski background, that is, setting $G=0$, and the initial metric to flat space. In doing so we can see how gravity supports the collapse, in addition to the non-linear self-interactions.

\section{Results of the Simulations}
\label{sec:Results}

In this Section we present the results of our NR simulations, showing the dynamics of the central field amplitude $\phi$ and of the metric conformal factor $\chi$. Generically for all the models, as can be seen in the figures,  in the metastable phase the central field amplitude oscillates in proper time such as in Eq.~\eqref{eq:ParametrizedProfile}, with $\Phi(\tau) \propto \cos(\omega \tau + \delta)$, $\omega = c\, m$ and $c=\mathcal{O}(1)$. The decrease of $\chi$ to zero at the center signals the formation of a black hole, that we check by finding an apparent horizon whose area gives a lower bound on the mass of the final black hole. For all the cases in which a collapse occurs, the final mass of the black hole, computed using the apparent horizon, is between $85\%$ and $95\%$ of the original mass of the RSS.

\subsection{$\alpha$-attractor T-models}

\begin{figure}[h!]
\centering
	\includegraphics[width=0.49\textwidth]{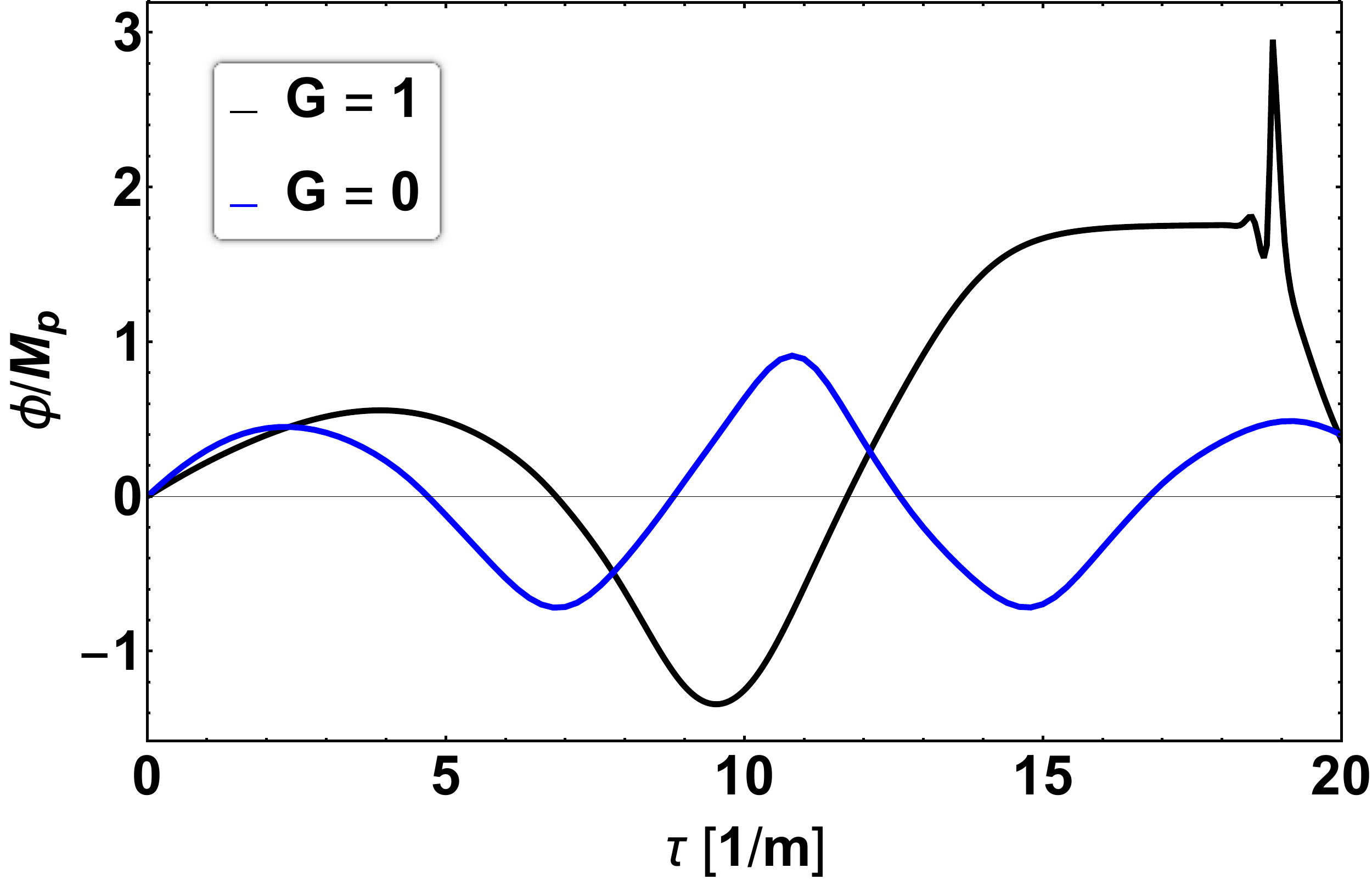}
    \includegraphics[width=0.49\textwidth]{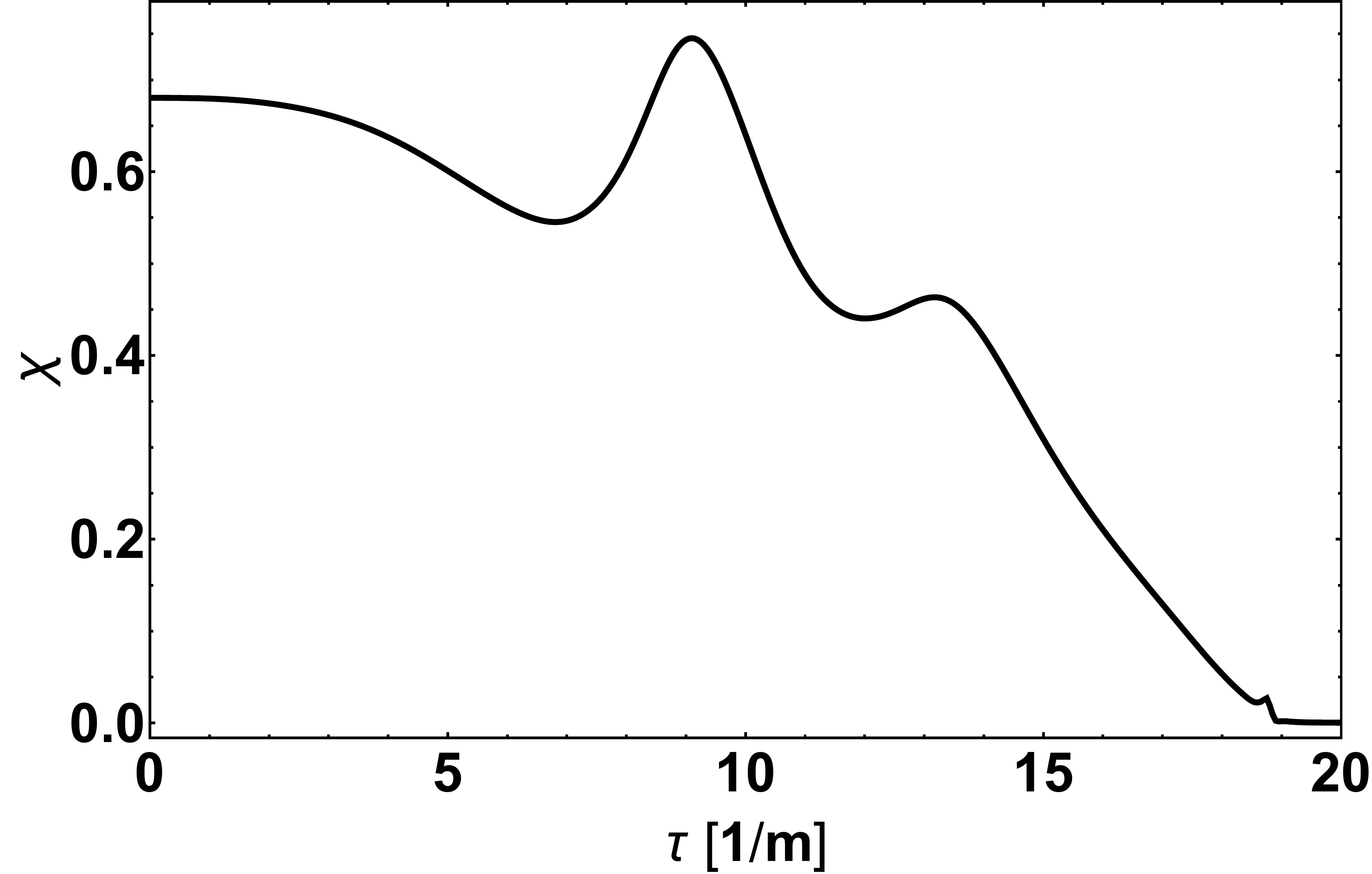}
    \caption{
    {\it T-model potential: } \textit{(Left panel)} 
   Dynamical evolution of the central amplitude of the oscillaton for a configuration with total initial mass $M = 2.9 \, M_p^2/m$ and typical scale of the potential $\Lambda/M_p = 0.2$. The black curve correspond to the evolution that takes into account the effects of gravity, while the blue curve is the stable evolution in the absence of gravity.
 The spike at $\tau \simeq 19 \times 1/m$ in the black curve is caused by the collapse to a black hole as the puncture gauge evolves, and is thus somewhat unphysical. The convergence test for this parameter point is shown in Fig.~\ref{fig:Converge}. 
 \textit{(Right panel)} Dynamical evolution of the metric conformal factor $\chi$ for $\Lambda/M_p = 0.2$ with initial total mass $M = 2.9 \, M_p^2/m$. The convergence test for this parameter point is shown in Fig.~\ref{fig:Converge}.}
\label{fig:phiT02phi03Comp}
\end{figure}

We consider values of the typical scale of the potential $0.14 < \Lambda/M_p < 0.2$ and we take for concreteness $n = 1$. These values are borderline for the formation of oscillons in the simplest scenario of single field with self-interactions, since the requirement for the formation in that case is $\Lambda \ll M_p$. For smaller values of the scale $\Lambda$, GR effects are always negligible, as already observed in~\cite{Amin:2019ums}. For each value of the typical scale of the potential we vary the initial total mass $M$ of the configuration in the range $2.6 \, M_p^2/m < M < 3.0 \, M_p^2/m$. We observe three peculiar behaviours for RSSs in T-model potentials:
\begin{enumerate}
\item For the smallest $\Lambda/M_p = 0.14$, all the solutions (corresponding to different initial total masses) radiate scalar waves during the first oscillations, settling down into an equilibrium configuration. We show this behaviour for the heaviest initial total mass $M = 3.0 \, M_p^2/m$ in Fig.~\ref{fig:phiT014phi04N32}. By the end of the simulation the RSS has lost $25\%$ of its initial total mass and the compactness has decreased by a factor $\mathcal{C}_{\rm initial}/\mathcal{C}_{\rm final} = 4.4$.

\begin{figure}[h!]
    \centering
    \includegraphics[width=0.85\textwidth]{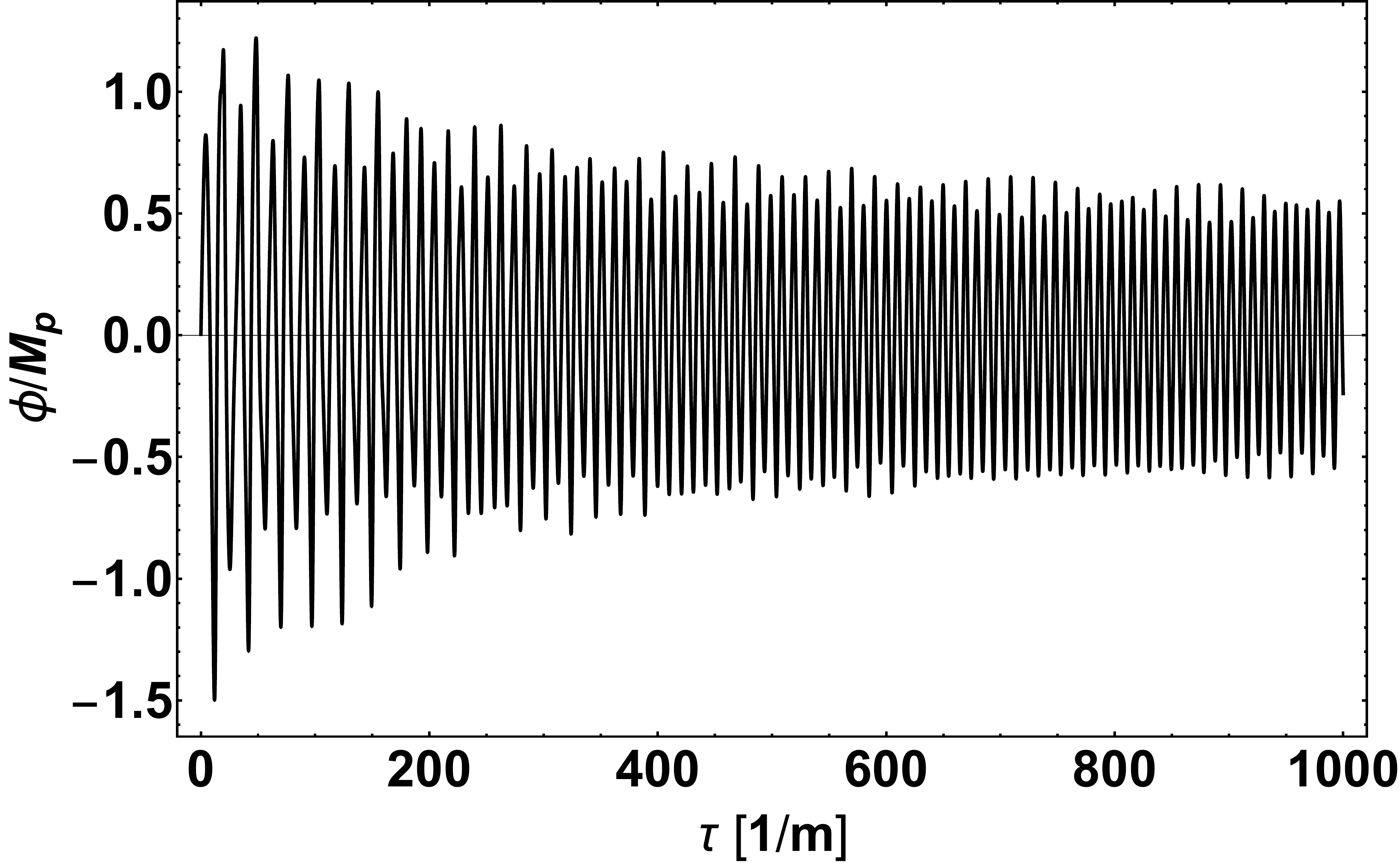}
    \caption{{\it T-model potential: }Evolution of the central field amplitude for the T-model potential with $n = 1$, $\Lambda/M_p = 0.14$ and initial total mass of the configuration $M = 3.0 \, M_p^2/m$. The RSS settles into a stable configuration after radiating part of the energy into scalar waves during the first oscillations.}
\label{fig:phiT014phi04N32}
\end{figure}

\item For $\Lambda/M_p = 0.17$ the configurations with initial total mass $2.9 \, M_p^2/m \leq M \leq 3.0 \, M_p^2/m$ collapse to black holes after a few oscillations with increasing amplitude, see left panel of Fig.~\ref{fig:T017phi0203} for the $M = 2.9 \, M_p^2/m$ case. The configuration with initial total mass $M = 2.6 \, M_p^2/m$ remains meta-stable after a couple of oscillations with large amplitude, during which it radiates part of the energy into scalar waves, see right panel of Fig.~\ref{fig:T017phi0203}.
\begin{figure}[h!]
    \centering
    \includegraphics[width=0.49\textwidth]{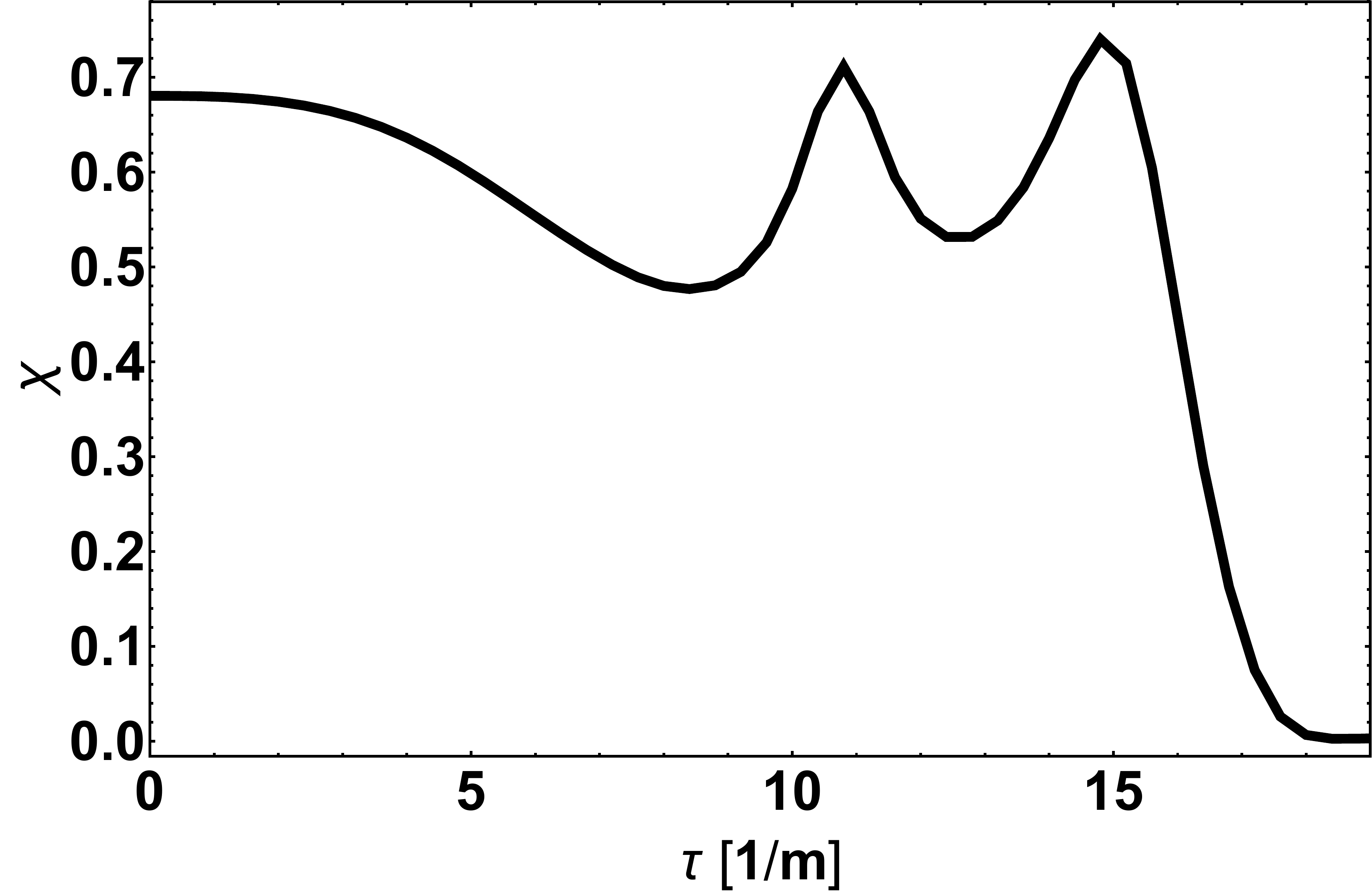}
    \includegraphics[width=0.5\textwidth]{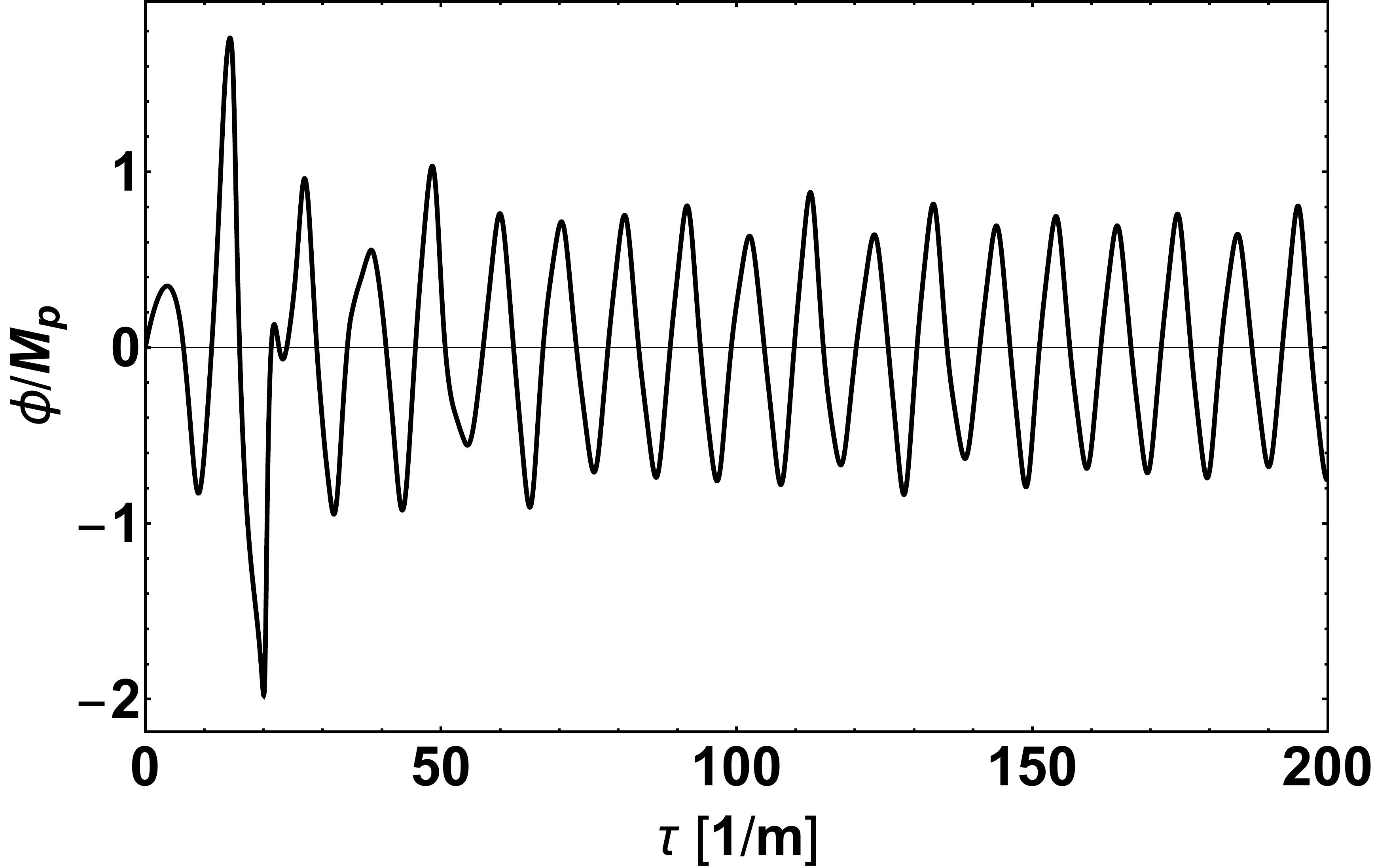}
    \caption{{\it T-model potential: }\textit{(Left panel)} Dynamical evolution of the conformal factor $\chi$ for the configuration with initial total mass $M = 2.9 \, M_p^2/m$ and $\Lambda/M_p = 0.17$. The configuration quickly collapses to a black hole. \textit{(Right panel)} Dynamical evolution of the central field amplitude $\phi$ for the configuration with initial mass $M = 2.6 \, M_p^2/m$ and $\Lambda/M_p = 0.17$. The configuration settles into an equilibrium configuration after radiating away part of the energy through scalar waves during the first two oscillations.}
\label{fig:T017phi0203}
\end{figure}
\item For $\Lambda/M_p = 0.2$ the behaviour is analogous to the case $\Lambda/M_p = 0.17$: configurations with initial total mass $2.9 \, M_p^2/m \leq M \leq 3.0 \, M_p^2/m$ collapse very rapidly to black holes, see Fig.~\ref{fig:phiT02phi03Comp}. Configurations with initial total mass $M = 2.6 \, M_p^2/m$ settles into a stable solution after radiating part of the energy into scalar waves during the first transient stage.
\end{enumerate}

\begin{figure}[h!]
    \centering
    \includegraphics[width=0.85\textwidth]{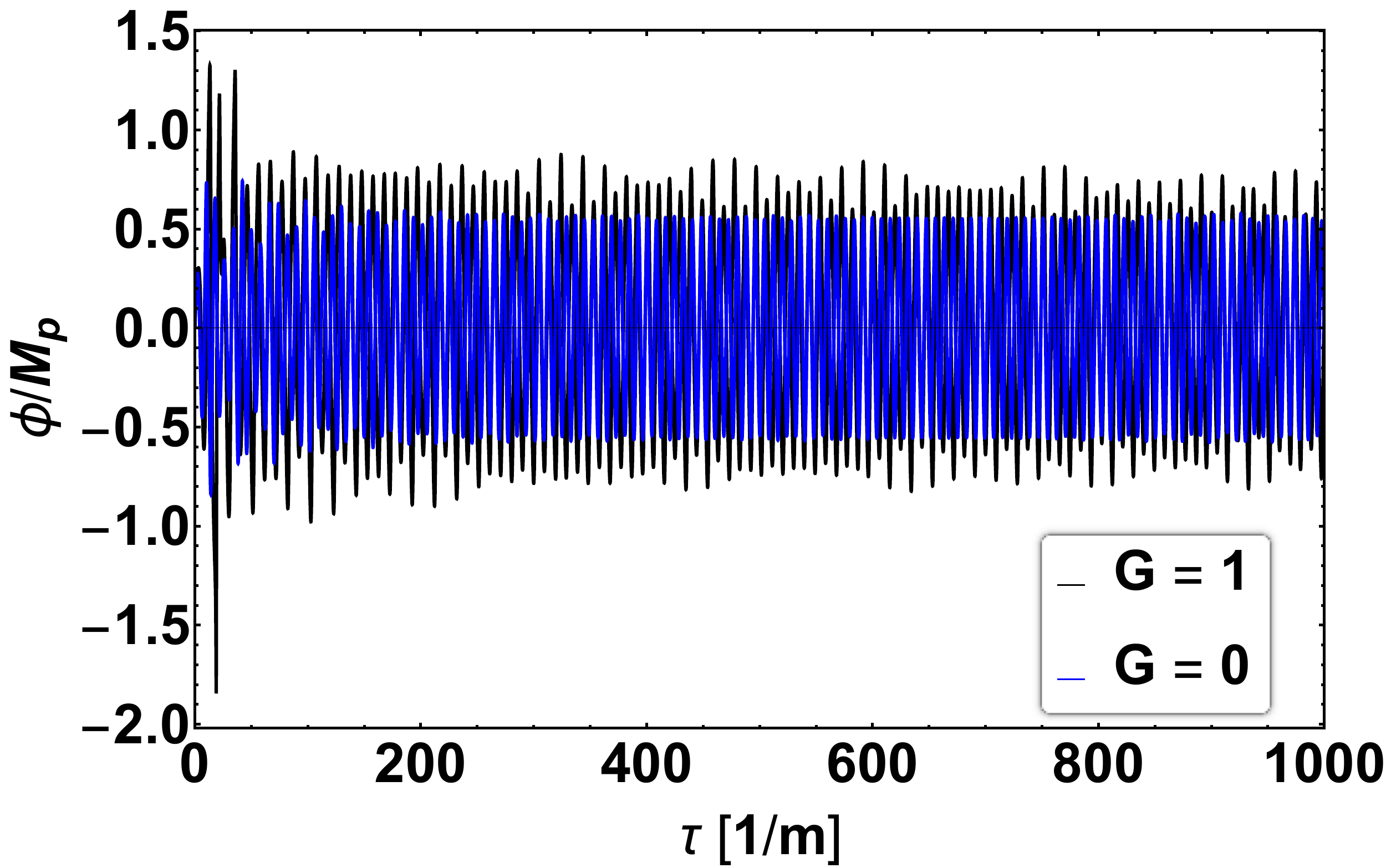}
    \caption{{\it T-model potential: }The black curve corresponds to the dynamical evolution of the central amplitude taking into account the effects of gravity, while the blue curve corresponds to the evolution of the same initial configuration neglecting gravity effects. The total initial mass is $M = 2.6 \, M_p^2/m$ and the typical scale of the potential is $\Lambda/M_p = 0.2$. Both solutions are stable - the latter due only to self-interactions - but the former has a larger amplitude of oscillations that could give rise to a larger stochastic background of GWs, as explained in Sec.~\ref{sec:Introduction} (see Eq.~\eqref{eq:GWSpectrumAmplitude}).}
\label{fig:phiT02phi02Comp}
\end{figure}

In Fig.~\ref{fig:phiT02phi02Comp} we show the comparison between the dynamical evolution of the central amplitude of two RSSs with the same initial total mass $M = 2.6 \, M_p^2/m$ when gravity is taken into account (black line) and when gravity are neglected (blue line), for $\Lambda/M_p = 0.2$. Upon adding the effects of gravity the equilibrium amplitude is larger: this could be an important effect for GW production in the light of Eq.~\eqref{eq:GWSpectrumAmplitude}, in the non-spherically symmetric cases. This effect is even more pronounced for stable RSSs with larger initial amplitude but not large enough to make them collapse to black holes.

\subsection{$\alpha$-attractor E-models}

\begin{figure}[h!]
    \centering
    \includegraphics[width=0.5\textwidth]{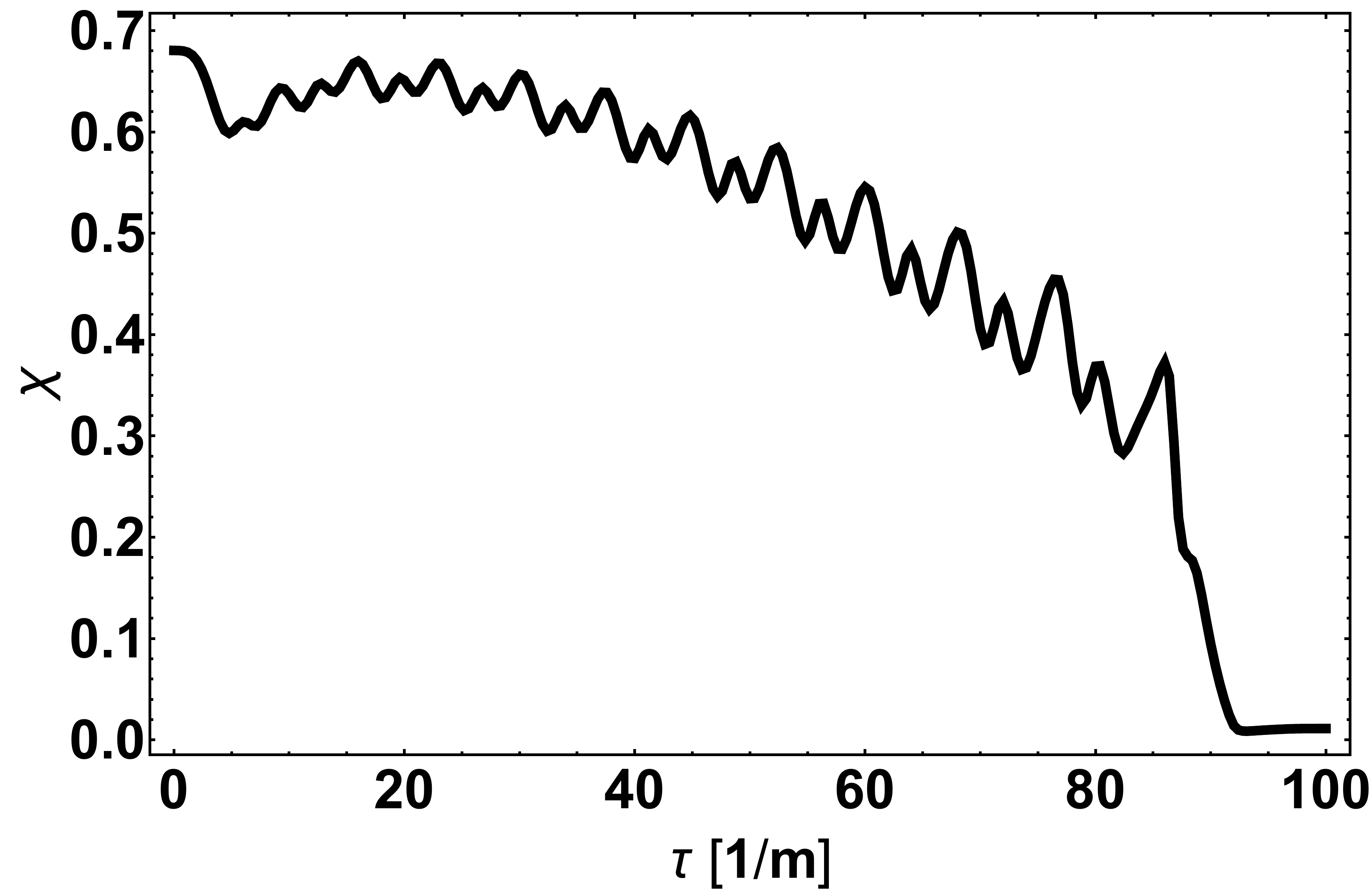}
    \includegraphics[width=0.49\textwidth]{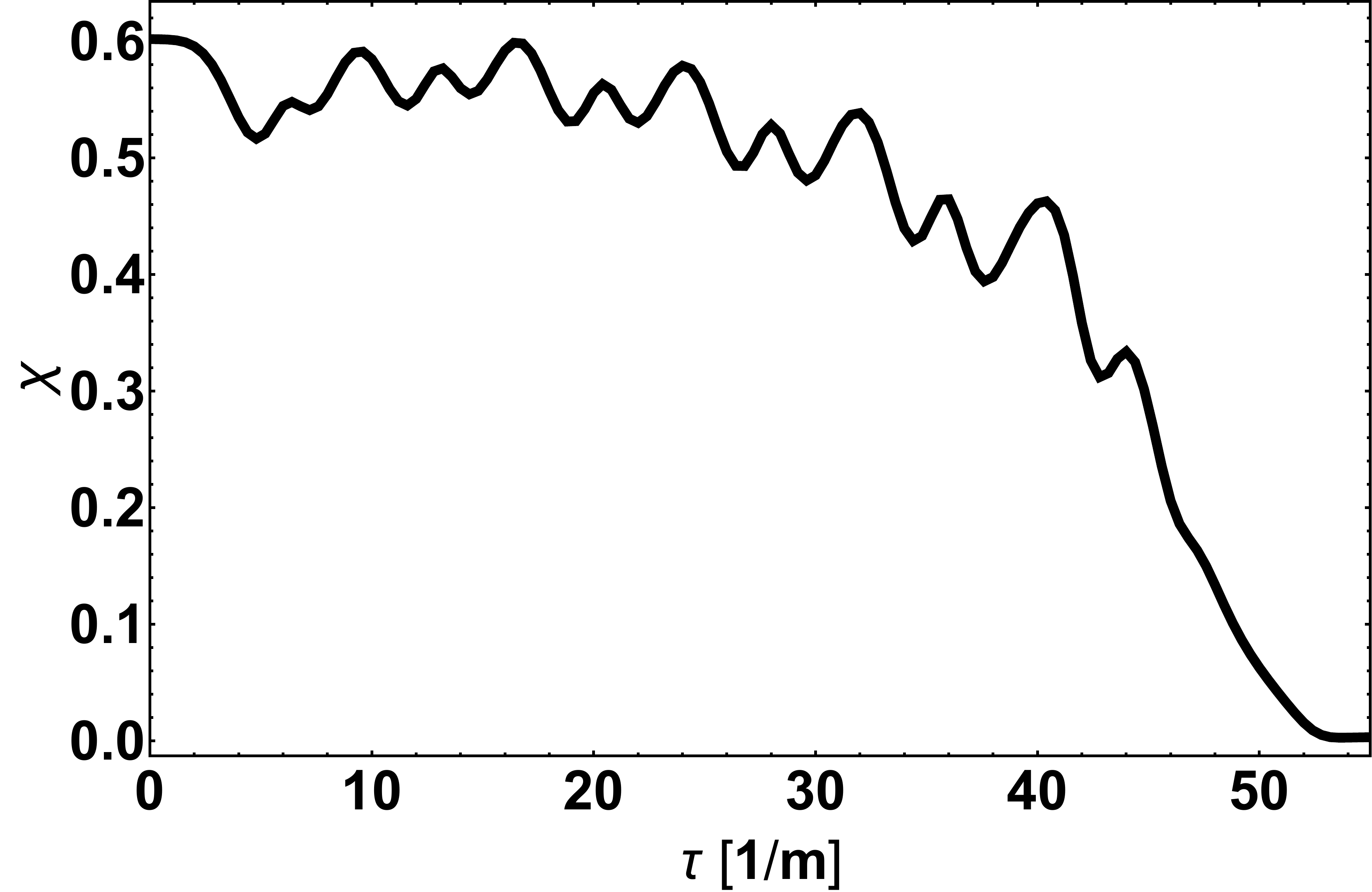}
    \caption{{\it E-model potential: } \textit{(Left panel)} Plot of the conformal factor $\chi$ in the case of an E-model potential with parameter $\Lambda/M_p = 1$ and initial total mass $M = 2.9 \, M_p^2/m$. The number of points in the coarsest grid for the simulation from which the curve is extracted is $N=96$. \textit{(Right panel)} Plot of the conformal factor $\chi$ in the case of an E-model potential with parameter $\Lambda/M_p = 1$ and initial total mass $M = 3.0 \, M_p^2/m$.}
\label{fig:chiSa1phi0304}
\end{figure}

\begin{figure}[h!]
    \centering
    \includegraphics[width=0.48\textwidth]{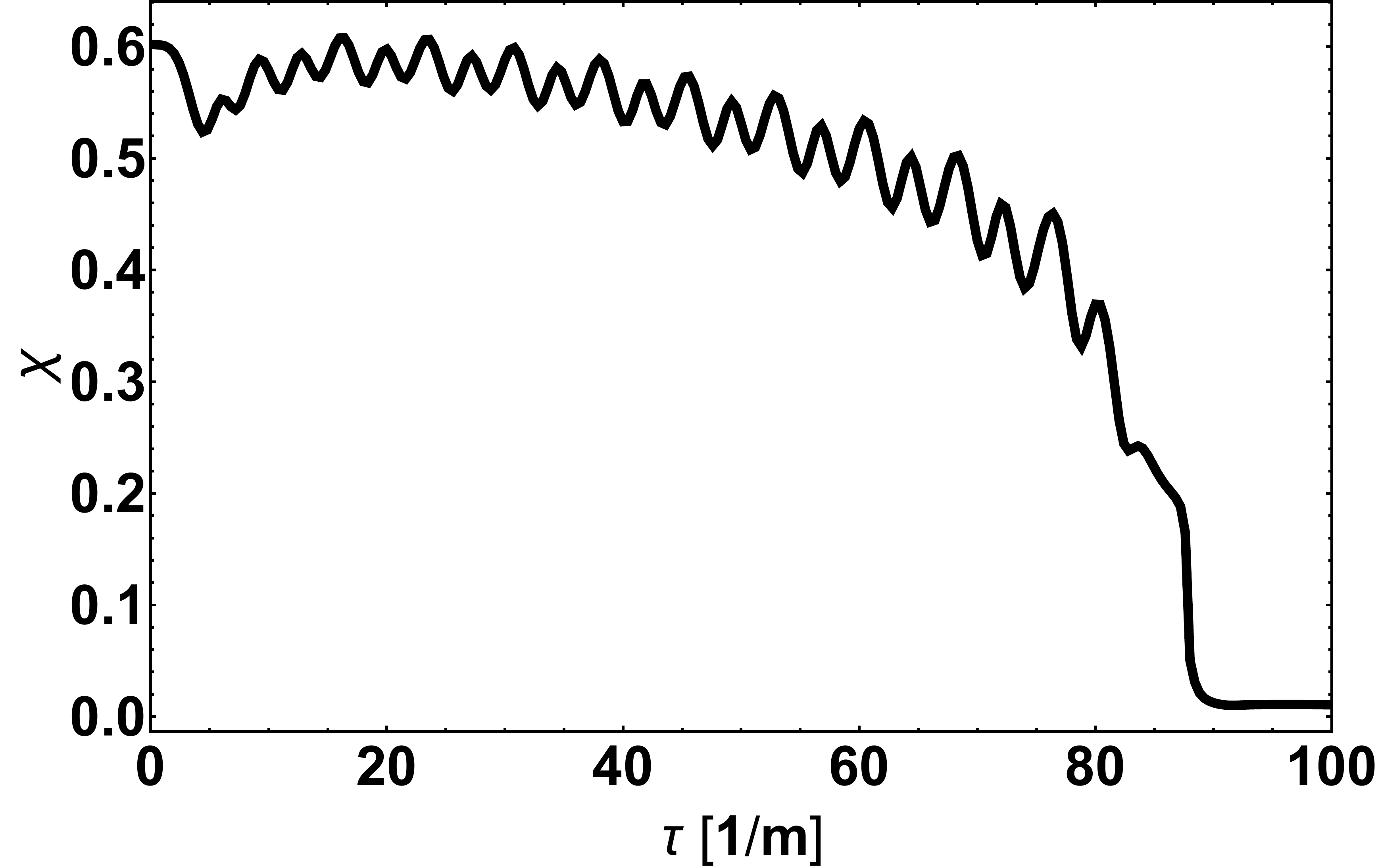}
    \includegraphics[width=0.49\textwidth]{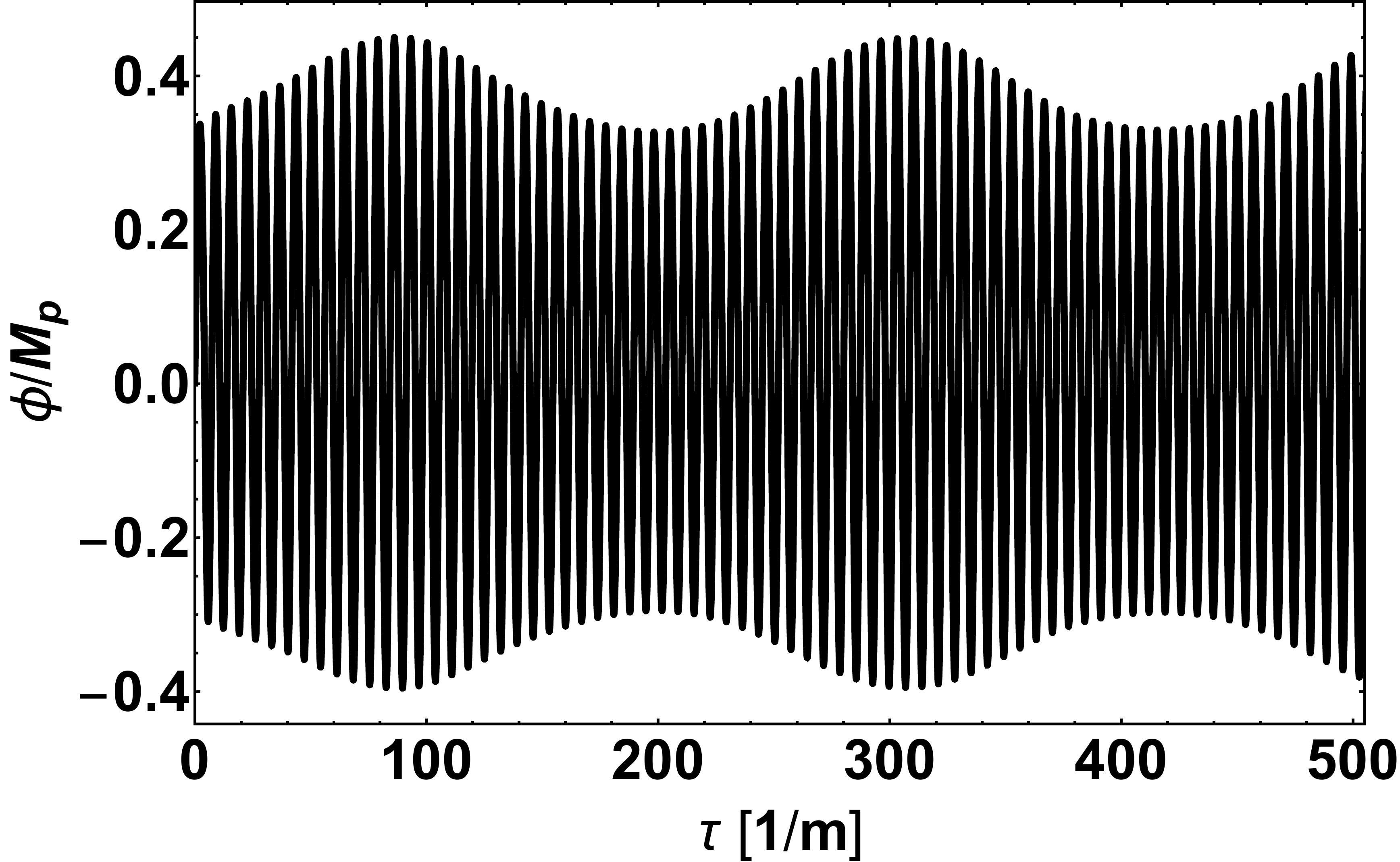}
    \caption{{\it E-model potential: }\textit{(Left panel)} Plot of the conformal factor $\chi$ in the case of an E-model potential with parameter $\Lambda/M_p = 2$ and initial total mass $M = 3.0 \, M_p^2/m$. \textit{(Right panel)} Central field profile $\phi$ in the case of an E-model potential with parameter $\Lambda/M_p = 2$ and initial total mass $M = 2.9 \, M_p^2/m$.}
\label{fig:chiSa2phi0304}
\end{figure}

In this case we consider two cases: $\Lambda/M_p = 1$ and $\Lambda/M_p = 2$, so as not to restrict ourselves to a single model. We observe the following behaviours:	\\
\begin{enumerate}
\item For $\Lambda/M_p = 1$ the configurations with initial total mass $2.9 \, M_p^2/m \leq M \leq 3.0 \, M_p^2/m$ collapse rapidly to black holes, see Fig.~\ref{fig:chiSa1phi0304}. The configuration with initial total mass $M = 2.6 \, M_p^2/m$ settles into a stable solution after radiating part of the energy into scalar waves.
\item For $\Lambda/M_p = 2$ the configuration with initial total mass $M = 3.0 \, M_p^2/m$ collapses to a black hole, see left panel of Fig.~\ref{fig:chiSa2phi0304}, while the configurations with initial total mass $2.6 \, M_p^2/m \leq M \leq 2.9 \, M_p^2/m$ settle into stable solutions after radiating part of the energy into scalar waves, see right panel of Fig.~\ref{fig:chiSa2phi0304}.
\end{enumerate}

\begin{figure}[h!]
    \centering
    \includegraphics[width=0.49\textwidth]{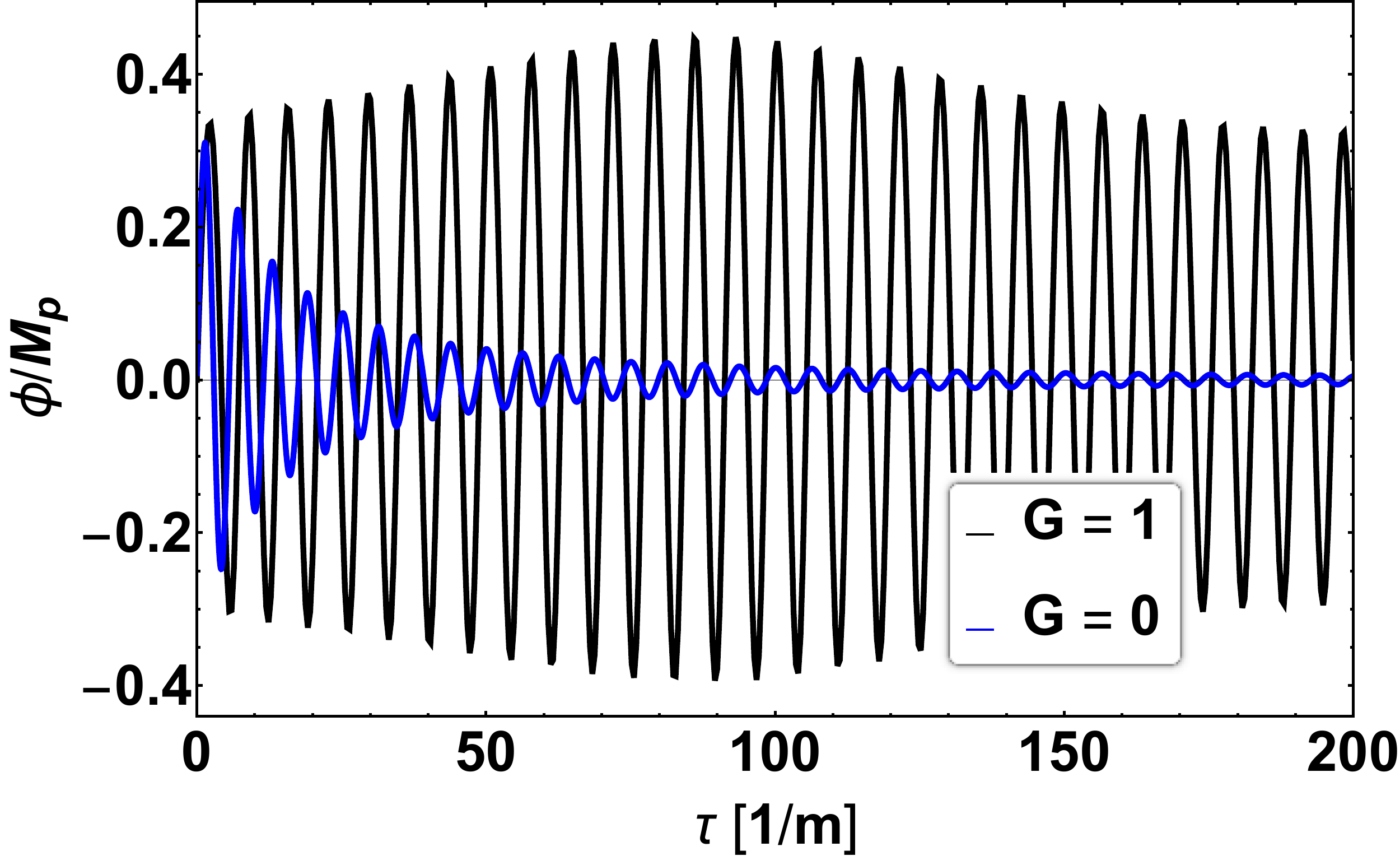}
    \includegraphics[width=0.49\textwidth]{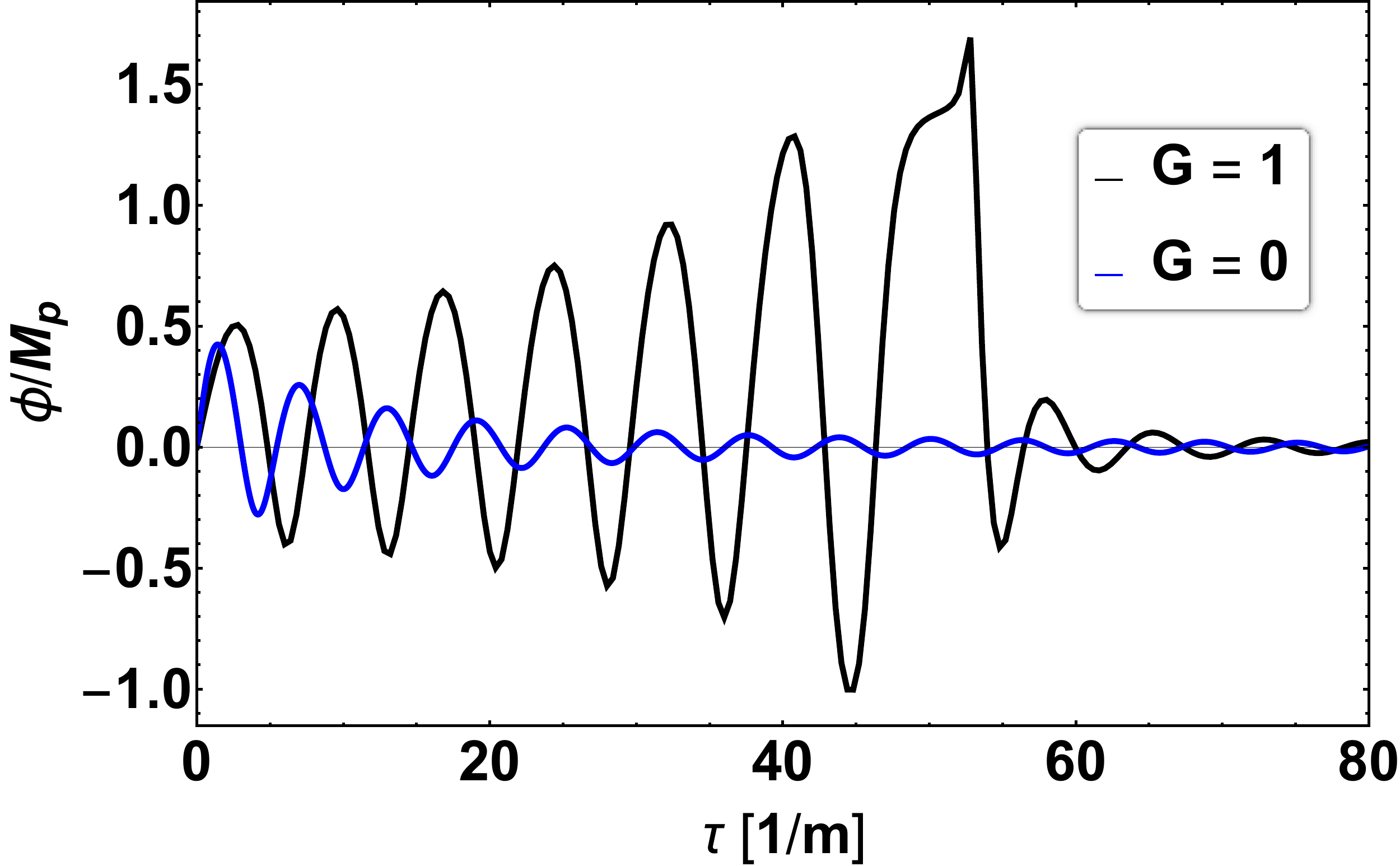}
    \caption{{\it E-model potential: }\textit{(Left panel)} Comparison between the dynamical evolution of the same initial configuration (total mass $M = 2.9 \, M_p^2/m$) in the E-model potential with $\Lambda/M_p = 2$ with and without gravity. The presence of gravity stabilizes the RSS even if the attractive self-interactions are not strong enough to make the configuration stable. \textit{(Right panel)} Comparison between the dynamical evolution of the same initial configuration (total mass $M = 3.0 \, M_p^2/m$) in the E-model potential with $\Lambda/M_p = 1$ with (black line) and without gravity (blue line). The presence of gravity drives the RSS to collapse to a black hole.}
\label{fig:phiSa2phi03Comp}
\end{figure}

In the left panel of Fig.~\ref{fig:phiSa2phi03Comp} we show the comparison between the stable configuration obtained including gravity effects in the simulation (black line) and the configuration with the same initial conditions if gravity effects are switched off (blue line): the presence of gravity makes the RSS stable even in the case in which the attractive self-interactions are not strong enough to allow the existence of the oscillon solution, as evident from the fact that the central field is damped very quickly, i.e. the RSS disperses. At the same time, in the right panel we show how the attractive gravitational interaction can dynamically drive the RSS to black hole collapse, even in a case that would quickly disperse in the absence of gravity effects.

\subsection{KKLT}

\begin{figure}[h!]
    \begin{center}
    \includegraphics[width=0.85\textwidth]{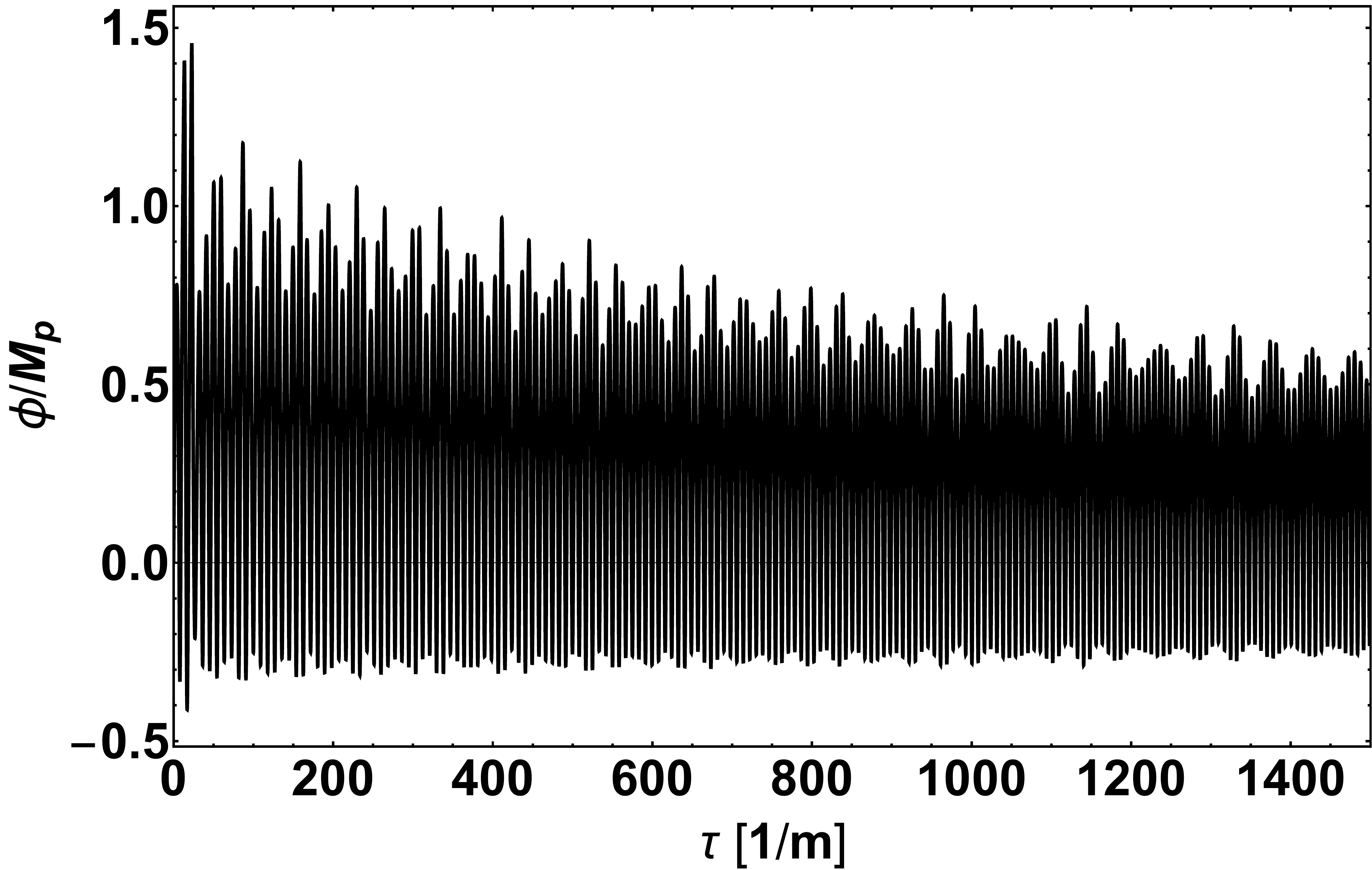}
    \end{center}
    \caption{{\it KKLT potential: } Evolution of the central amplitude for the KKLT potential with $|W_0| = 10^{-2}$ and initial total mass $M = 3.0 \, M_p^2/m$. The RSS slowly radiates energy into scalar waves until it settles into an equilibrium configuration. At the very end of the simulation the RSS seems to be very slowly losing energy through scalar radiation. Likely, it will keep losing energy until it reaches an equilibrium configuration close to that obtained by setting $G = 0$, see Fig.~\ref{fig:phiW02phi04Comp}.}
\label{fig:phiW02phi04}
\end{figure}

\begin{figure}[h!]
    \centering
    \includegraphics[width=0.85\textwidth]{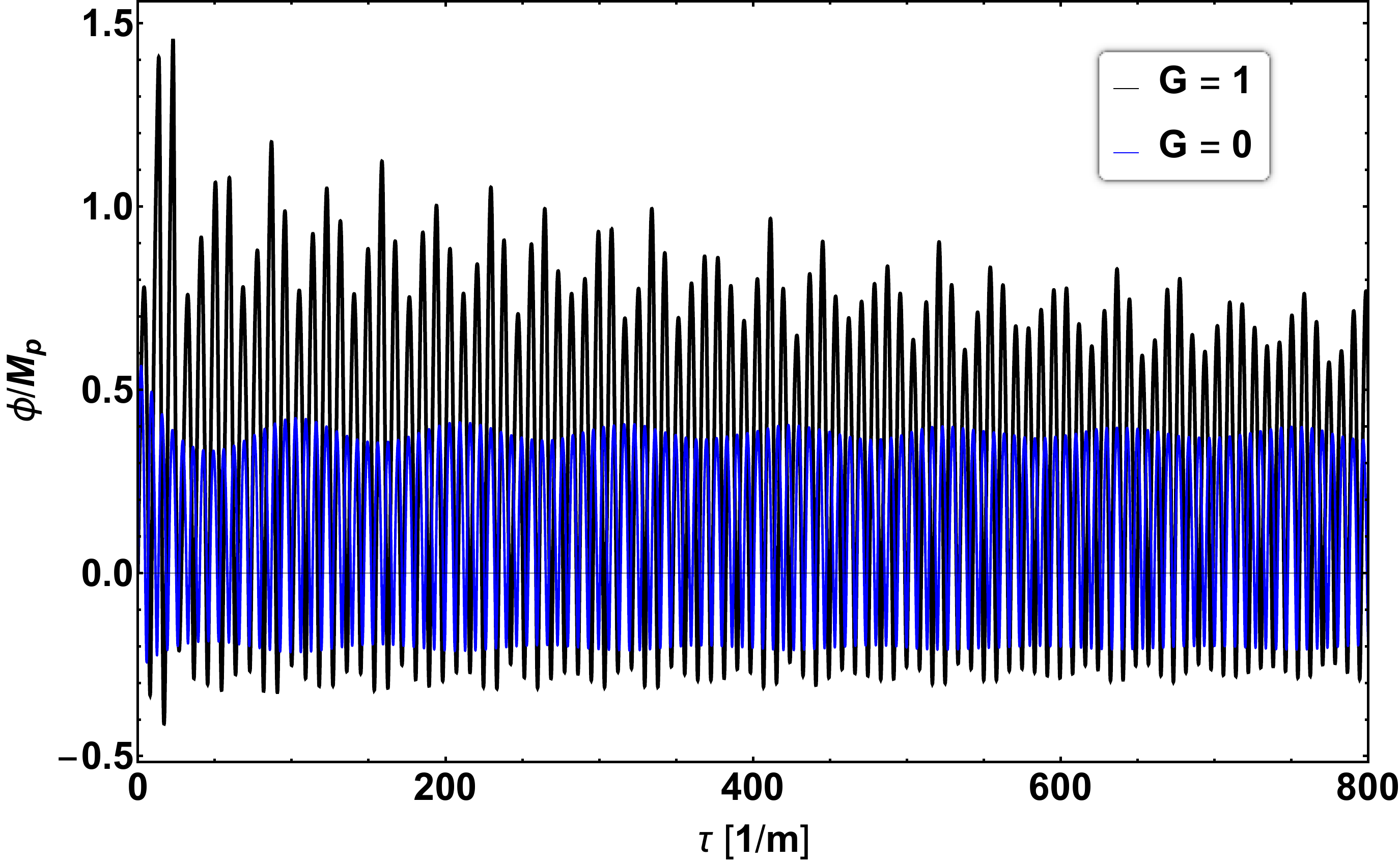}
    \caption{{\it KKLT potential: }Comparison between the evolutions of the central amplitude for the KKLT potential with $|W_0| = 10^{-2}$ including the effects of gravity ($G = 1$) and neglecting them $(G = 0)$. The initial total mass is $M = 3.0 \, M_p^2/m$. In the latter case the RSS radiates the excess energy into scalar waves very quickly, until it settles into an equilibrium configuration. In the former case the decay is very slow and the oscillation amplitude is much larger for a long time.}
\label{fig:phiW02phi04Comp}
\end{figure}

In the KKLT scenario we observe the same qualitative behaviour for all the initial total masses in the range $2.6 \, M_p^2/m < M < 3.0 \, M_p^2/m$: the RSS radiates part of the energy into scalar waves until it settles into a quasi-equilibrium configuration. We never observe collapse to black holes in this scenario for the region of the parameter space investigated (i.e. $|W_0| \gtrsim 10^{-4}$) for which the formation of oscillons has been studied in~\cite{Antusch:2017flz}. In Fig.~\ref{fig:phiW02phi04} we show the evolution of the central field amplitude for the parameter point $|W_0| = 10^{-2}$ and initial total mass $M = 3.0 \, M_p^2/m$. This point has been chosen in such a way to maximize the typical scale of the potential. At the very end of the simulation the RSS has lost $44\%$ of its mass and it seems to be still radiating energy into scalar waves very slowly. The compactness is reduced by a factor $\mathcal{C}_{\rm initial}/\mathcal{C}_{\rm final} \simeq 5$. Notice the remarkably different behaviour with respect to the case without gravity shown in Fig.~\ref{fig:phiW02phi04Comp}: the oscillation amplitude is initially much larger if the effects of gravity are included. 

\section{Conclusions}
\label{sec:Conclusions}

We have analyzed by means of full GR simulations the stability of RSS solutions in various potentials relevant for inflation and for realistic string theory scenarios. Assuming large relativistic perturbations of various masses as an initial condition, we have used NR techniques to study their dynamical evolution and stability. The results are summarized in Tab.~\ref{tab:Summary}, including the results obtained in the absence of gravity effects. The present analysis is relevant for a better understanding of the post-inflationary dynamics and the possible production of GWs during the pre-BBN era. We have observed several different behaviours, depending on the potential, the parameters of the model and the mass of the initial RSS configuration. While we observe that in KKLT scenarios RSSs are meta-stable in the region of parameter space investigated even for the heaviest initial configurations taken into account, for the other potentials the heaviest RSSs can be dynamically driven (by the interplay of gravity and self-interactions) to collapse to black holes, even if the initial radii are much larger than the corresponding Schwarzschild radii. We have performed a comparison with the corresponding solutions that would be obtained in the absence of gravity, showing that gravitational effects can significantly modify the dynamics of the RSSs in the region of parameter space taken into account. In particular we observe the following general features:
\begin{itemize}
\item In cases in which the RSS quickly disperses in the absence of gravity, gravitational effects can either stabilize it, or drive it to collapse to black hole, see e.g. Fig.~\ref{fig:phiSa2phi03Comp} for the case with E-model potential. Hence, one would expect to observe more RSSs (and hence enhanced GW production) than those predicted by lattice simulations of preheating scenarios that do not take into account gravity. Further studies to better understand  the formation of RSSs in these potentials are required, to determine under what circumstances perturbations can grow to relativistic amplitudes.
\item In some cases in which the RSS is stable with self-interactions only, gravitational effects can either enhance the amplitude of the field oscillations in its interior, see e.g. Fig.~\ref{fig:phiT02phi02Comp} (T-models) and Fig.~\ref{fig:phiW02phi04Comp} (KKLT), or dynamically drive it to collapse to black holes, see e.g. left panel of Fig.~\ref{fig:phiT02phi03Comp} (T-models). While the former effect can have a significant impact on the production of GWs due to the RSS dynamics, as explained in Sec.~\ref{sec:Introduction} (see Eq.~\eqref{eq:GWSpectrumAmplitude}), the latter effect would significantly affect the reheating history of the Universe.
\end{itemize}

Numerical simulations are valuable tools for understanding the physics of the very early Universe. The simulations we have performed can be easily adapted to study more complex systems of perturbations in the absence of spherical symmetry. Some possible future directions include the production of relativistic oscillatons during preheating, the formation of primordial black holes, and the production of GWs from early matter domination. We plan to return to these and related questions in future work.\\

\begin{table}[h!]
\centering
\begin{tabular}{|c|c|c|c|c|c|c|c|}
\hline
\multirow{2}{*}{Model} & \multirow{2}{*}{Parameters} & \multicolumn{2}{c|}{$M = 2.6 \, M_p^2/m$} & \multicolumn{2}{c|}{$M = 2.9 \, M_p^2/m$} & \multicolumn{2}{c|}{$M = 3.0 \, M_p^2/m$} \\ \cline{3-8} 
                  &                   & $G = 0$ & $G = 1$ & $G = 0$ & $G = 1$ & $G = 0$ & $G = 1$ \\ \hline \hline
Free-Field & $\times$ & D & MS & D & MS & D & MS \\ \hline \hline 
\multirow{3}{*}{} & $\Lambda/M_p = 0.14$ & MS & MS & MS & MS &          MS & MS \\ \cline{2-8} 
T-models & $\Lambda/M_p = 0.17$ & MS & MS & MS &         C & MS & C \\ \cline{2-8} 
                  & $\Lambda/M_p = 0.2$ & MS & MS & MS & C & MS &          C \\ \hline \hline
\multirow{2}{5em}{E-models} & $\Lambda/M_p = 1$ & D &          MS & D & C & D & C \\ \cline{2-8} 
                  & $\Lambda/M_p = 2$ & D &  MS & D & MS & D & C \\ \hline \hline
KKLT & $|W_0| = 10^{-2}$ & MS & MS & MS & MS & MS & MS \\ \hline
\end{tabular}
\caption{Summary of the results. The capital letters stand for: D = Dispersion, MS = Meta-Stable, C = Collapse to a black hole. For completeness, we show the comparison with the free-field case: the stability in that case is achieved only if the configuration lies on the stability curve, as explained in~\cite{UrenaLopez:2001tw}. \label{tab:Summary}}
\end{table}

\section*{Acknowledgments}

We appreciate interesting discussions with Francesco Cefal\'a and Stefano Orani. We thank Ricardo Becerril and Thomas Helfer for having provided the initial conditions for stable oscillatons. KC acknowledges support from the European Research Council. Our simulations used the Argo cluster at ICTP and we thank Ivan Girotto for his support with using the system. We acknowledge CINECA, for the availability of high performance computing resources and support which were also used in this work. Finally some simulations were performed using the Cambridge Service for Data Driven Discovery (CSD3), part of which is operated by the University of Cambridge Research Computing on behalf of the STFC DiRAC HPC Facility (www.dirac.ac.uk). The DiRAC component of CSD3 was funded by BEIS capital funding via STFC capital grants ST/P002307/1 and ST/R002452/1 and STFC operations grant ST/R00689X/1. DiRAC is part of the UK National eInfrastructure.

\end{document}